\documentclass[12pt]{article}
\usepackage{arxiv}
\usepackage[utf8]{inputenc} % allow utf-8 input
\usepackage[T1]{fontenc} 
\usepackage{graphicx} % load for figures
\usepackage{caption}
\usepackage{subcaption}
\usepackage{amsthm, bm}
\usepackage{amsfonts} % load for \mathbb
\usepackage{amsmath}  % load for \boldsymbol
\usepackage{float}
\usepackage{comment}
\usepackage{xcolor}
\usepackage{xr}
\usepackage{hyperref} % for cross-reference
\let\bibhang\relax
\usepackage{natbib}   % load for reference

\makeatletter
\newcommand*{\addFileDependency}[1]{% argument=file name and extension
	\typeout{(#1)}
	\@addtofilelist{#1}
	\IfFileExists{#1}{}{\typeout{No file #1.}}
}
\makeatother

\newcommand*{\myexternaldocument}[1]{%
	\externaldocument{#1}%
	\addFileDependency{#1.tex}%
	\addFileDependency{#1.aux}%
}

\usepackage[bottom]{footmisc}   % fix the footnote of tables

\myexternaldocument{Web_Appendix}

\newcommand{\given}{\,|\,}

\newcommand{\bbeta}{\boldsymbol{\beta}}

\newcommand{\bzero}{\mathbf{0}}
\newcommand{\calE}{{\cal E}}

\newcommand{\calG}{{\cal G}}
\newcommand{\calV}{{\cal V}}
\newcommand\myeq{\mathrel{\overset{\makebox[0pt]{\mbox{\normalfont\tiny\sffamily ind}}}{\sim}}}

\title{Hierarchical Multivariate Directed Acyclic
	Graph Auto-Regressive (MDAGAR) models for
	spatial diseases mapping}

\author{Leiwen Gao\\
	Department of Biostatistics, University of California, Los Angeles\\
	\texttt{gaoleiwen@ucla.edu} \\ \\
	\and
	\textbf{Abhirup Datta}\\
	Department of Biostatistics, Johns Hopkins University\\
	\texttt{abhidatta@jhu.edu} \\ \\
	\and
	\textbf{Sudipto Banerjee}\\
	Department of Biostatistics, University of California, Los Angeles\\
	\texttt{sudipto@ucla.edu} 
}

\begin{document}

	\maketitle
	
\begin{abstract}
	Disease mapping is an important statistical tool used by epidemiologists to assess geographic variation in disease rates and identify lurking environmental risk factors from spatial patterns. Such maps rely upon spatial models for regionally aggregated data, where neighboring regions tend to exhibit similar outcomes than those farther apart. We contribute to the literature on multivariate disease mapping, which deals with measurements on multiple (two or more) diseases in each region. We aim to disentangle associations among the multiple diseases from spatial autocorrelation in each disease. We develop Multivariate Directed Acyclic Graphical Autoregression (MDAGAR) models to accommodate spatial and inter-disease dependence. The hierarchical construction imparts flexibility and richness, interpretability of spatial autocorrelation and inter-disease relationships, and computational ease, but depends upon the order in which the cancers are modeled. To obviate this, we demonstrate how Bayesian model selection and averaging across orders are easily achieved using bridge sampling. We compare our method with a competitor using simulation studies and present an application to multiple cancer mapping using data from the Surveillance, Epidemiology, and End Results (SEER) Program.
\end{abstract}


\maketitle

\section{Introduction}
Spatially-referenced data comprising regional aggregates of health outcomes over delineated administrative units such as counties or zip codes are widely used by epidemiologists to map mortality or morbidity rates and better understand their geographic variation. Disease mapping, as this exercise is customarily called, employs statistical models to present smoothed maps of rates or counts of a disease. Such maps can assist investigators in identifying lurking risk factors \citep{koch2005cartographies} and in detecting ``hot-spots'' or spatial clusters emerging from common environmental and socio-demographic effects shared by neighboring regions. By interpolating estimates of health outcome from areal data onto a continous surface, disease mapping also generates smoothed maps for the small-area scale, adjusting for the sparsity of data or low population size \citep{berke2004exploratory, richardson2004interpreting}. 

For a single disease, there has been a long tradition of employing Markov random fields (MRFs) \citep{rueheld04} to introduce conditional dependence for the outcome in a region given its neighbors. Two conspicuous examples are the Conditional Autoregression (CAR) \citep{besag1974spatial, besag1991bayesian} and Simultaneous Autoregression (SAR) models \citep{kissling2008spatial} that build dependence using undirected graphs to model geographic maps. More recently, \cite{datta2018spatial} proposed a class of Directed Acyclic Graphical Autoregressive (DAGAR) models as a preferred alternative to CAR or SAR models in allowing better identifiability and interpretation of spatial autocorrelation parameters. 

Multivariate disease mapping is concerned with the analysis of multiple diseases that are associated among themselves and across space. It is not uncommon to find substantial associations among different diseases sharing genetic and environmental risk factors. Quantification of genetic correlations among multiple cancers have revealed associations among several cancers including lung, breast, colorectal, ovarian and pancreatic cancers \citep{lindstrom2017quantifying}. Disease mapping exercises with lung and esophageal cancers have also evinced associations among them \citep{jin2005generalized}. When the diseases are inherently related so that the prevalence of one in a region encourages (or inhibits) occurrence of the other on the same unit, there can be substantial inferential benefits in jointly modeling the diseases rather than fitting independent univariate models for each disease \citep[see, e.g.,][]{knorr2001shared, kim2001bivariate, gelfand2003proper, carlin2003hierarchical, held2005towards, jin2005generalized, jin2007order, zhang2009smoothed, diva2008parametric, martinez2013general, mari2014smoothed}. 

Broadly speaking, there are two approaches to multivariate areal modeling. One approach builds upon a linear transformations of latent effects \citep[see, e.g.,][]{gelfand2003proper, carlin2003hierarchical,jin2005generalized,zhu2005, martinez2013general,bradley2015}. A different class emerges from hierarchical constructions \citep{jin2005generalized,daniels2006} where each disease enters the model in a given sequence. Here, we build a class of multivariate DAGAR (MDAGAR) models for multiple disease mapping by building the joint distribution hierarchically using univariate DAGAR models. 
We build upon the idea in \cite{jin2005generalized} of constructing GMCAR models, but with some important modifications. As noted in \cite{jin2005generalized}, the order in which the new diseases enter the hierarchical model specifies the joint distribution. Therefore, every ordering produces a different GMCAR model, which leads to an explosion in the number of models even for a modest number of cancers (say, more than 2 or 3 diseases). Alternatively, joint models that are invariant to ordering are constructed using linear transformations of latent random variables \citep{jin2007order}. However, these models are cumbersome and computationally onerous to fit and interpreting spatial autocorrelation becomes challenging.        

Our current methodological innovation lies in devising a hierarchical MDAGAR model in conjunction with {a bridge sampling algorithm \citep{meng1996simulating, gronau2017tutorial}} for choosing among differently ordered hierarchical models. The idea is to begin with a fixed ordered set of cancers, posited to be associated with each other and across space, and build a hierarchical model. The DAGAR specification produces a comprehensible association structure, while bridge sampling allows us to rank differently ordered models using their marginal posterior probabilities. Since each model corresponds to an assumed conditional dependence, the marginal posterior probabilities will indicate the tenability of such assumptions given the data. Epidemiologists, then, will be able to use this information to establish relationships among the diseases and spatial autocorrelation for each disease.         

The balance of this paper proceeds as follows. Section~\ref{Methods} develops the hierarchical MDAGAR model and {introduces a bridge sampling method} to select the MDAGAR with the best hierarchical order. Section~\ref{simulation} presents a simulation study to compare the MDAGAR with the GMCAR model and also illustrates the bridge sampling algorithm's efficacy in selecting the ``true'' model. Section~\ref{data} applies our MDAGAR to age-adjusted incidence rates of four cancers from the SEER database and discusses different cases with respect to predictors. Finally, in Section~\ref{discussion}, we summarize some concluding remarks and suggest promotion in the future research.
\section{Methods}
\label{Methods}
%	Our approach will be to construct a probability model of spatial random effects for each disease using the distribution specified by DAGAR, extending the univariate DAGAR to a multivariate model. We develop notations and briefly discuss the univariate DAGAR in the next section, followed by the multivariate extension in the following section. 

\subsection{Overview of Univariate DAGAR Modeling}\label{Univariate}
Let $\calG = \{\calV, \calE\}$ be a graph corresponding to a geographic map, where $\calV = \{1,2,\ldots,k\}$ is a fixed ordering of the vertices of the graph representing clearly delineated regions on the map, and $\calE = \{(i,j): i\sim j\}$ is the collection of edges between the vertices representing neighboring pairs of regions. We denote two neighboring regions by $\sim$. The DAGAR model, proposed by \citet{datta2018spatial}, builds a spatial autocorrelation model for a single outcome on $\calG$ using an ordered set of vertices in $\calV$. Let $N(1)$ be the empty set and let $N(j) = \{j' < j : j' \sim j\}$, where $j \in \calV \setminus \{1\}$. Thus, $N(j)$ includes geographic neighbors of region $j'$ that \emph{precede} $j$ in the ordered set $\calV$. Let $\{w_i : i\in \calV\}$ be a collection of $k$ random variables defined over the map. DAGAR specifies the following autoregression,
\begin{align}\label{eq:wj}
w_1 &= \epsilon_1; \quad w_j =  \sum_{j' \in N(j)} b_{jj'}w_{j'} + \epsilon_j,\; j = 2, \dots, k\;,
\end{align}
where $\epsilon_j \myeq N(0, \lambda_j)$ with the precision $\lambda_j$, and $b_{jj'}=0$ if $j' \not\in N(j)$. This implies that $\bm{w} \sim N(\bm{0}, \tau \bm{Q}(\rho))$, where $\bm{Q}(\rho)$ is a spatial precision matrix that depends only upon a spatial autocorrelation parameter $\rho$ and $\tau$ is a positive scale parameter. The precision matrix $\bm{Q}(\rho) = (\bm{I}-\bm{B})^{\top}\bm{F}(\bm{I}-\bm{B})$, $\bm{B}$ is a $k\times k$ strictly lower-triangular matrix and $\bm{F}$ is a $k\times k$ diagonal matrix. The elements of $\bm{B}$ and $\bm{F}$ are denoted by $b_{jj'}$ and $\lambda_{j}$, respectively, where 
\begin{align}\label{eq: B_and_F}
b_{jj'} = \left\{\begin{array}{l}
0\; \mbox{ if }\; j' \notin N(j)\;; \\
\frac{\rho}{1 + (n_{<j}-1)\rho^2}\; \mbox{ if }\; j=2,3,\ldots,k\;,\; j'\in N(j)\;	
\end{array}	 \right. \quad \mbox{and}\quad \lambda_{j} = \frac{1 + (n_{<j}-1)\rho^2}{1-\rho^2}\; j=1,2\ldots,k\;,
\end{align}
$n_{<j}$ is the number of members in $N(j)$ and $n_{<1} = 0$. The above definition of $b_{jj'}$ is consistent with the lower-triangular structure of $\bm{B}$ because $j'\notin N(j)$ for any $j'\geq j$. The derivation of $\bm{B}$ and $\bm{F}$ as functions of a spatial correlation parameter $\rho$ is based upon forming local autoregressive models on embedded spanning trees of subgraphs of $\calG$ \citep{datta2018spatial}. 

\subsection{Motivating multivariate disease mapping}\label{subsec: multivariate_disease_mapping}
There is a substantial literature on joint modeling of multiple spatially oriented outcomes, some of which have been cited in the Introduction. While it is possible to model each disease separately using a univariate DAGAR, hence independent of each other, the resulting inference will ignore the association among the diseases. This will be manifested in model assessment because the less dependence among diseases that a model accommodates, the farther away it will be from the joint model in the sense of Kullback-Leibler divergence.

More formally, suppose we have two mutually exclusive sets $A$ and $B$ that contain labels for diseases. Let $\bm{y}_A$ and $\bm{y}_B$ be the vectors of spatial outcomes over all regions corresponding to the diseases in set $A$ and set $B$, respectively. A full joint model $p(\bm{y})$, where $\bm{y} = \left(\bm{y}_A^{\top},\bm{y}_B^{\top}\right)^{\top}$, can be written as $p(\bm{y}) = p(\bm{y}_A)\times p(\bm{y}_B\given \bm{y}_A)$. Let $C_1$ and $C_2$ be two nested subsets of diseases in $A$ such that $C_2 \subset C_1 \subset A$. Consider two competing models,
$ p_1(\bm{y}) = p(\bm{y}_A)\times p(\bm{y}_B\given \bm{y}_{C_1})$ and $p_2(\bm{y}) = p(\bm{y}_A)\times p(\bm{y}_B\given \bm{y}_{C_2})$, where $p_1(\cdot)$ and $p_2(\cdot)$ are probability densities constructed from the joint probability measure $p(\cdot)$ by imposing conditional independence such that $p(\bm{y}_B\given \bm{y}_A) = p(\bm{y}_B\given \bm{y}_{C_1})$ and $p(\bm{y}_B\given \bm{y}_A) = p(\bm{y}_B\given \bm{y}_{C_2})$, respectively. Both $p_1(\cdot)$ and $p_2(\cdot)$ suppress dependence by shrinking the conditional set $A$, but $p_2(\cdot)$ suppresses more than $p_1(\cdot)$. We show below that $p_2(\cdot)$ is farther away from $p(\cdot)$ than $p_1(\cdot)$.

A straighforward application of Jensen's inequality yields $\displaystyle
\mathbb{E}_{B|C_1}\left[\log\frac{p(\bm{y}_{B}\given \bm{y}_{C_1})}{p(\bm{y}_{B}\given \bm{y}_{C_2})}\right] \geq 0$, where $\mathbb{E}_{B\given C_1}[\cdot]$ denotes the conditional expectation with respect to $p(\bm{y}_B\given \bm{y}_{C_1})$. Therefore,
\begin{equation}\label{eq: KL_div_mult_cancer}
\begin{split}
\mbox{KL}(p \| p_2) &- \mbox{KL}(p \| p_2) = \mathbb{E}_{A,B}\left[\log\left(\frac{p(\bm{y})}{p_2(\bm{y})}\right) - \log\left(\frac{p(\bm{y})}{p_2(\bm{y})}\right)\right]\\
&= \mathbb{E}_{A,B}\left[\log\frac{p_1(\bm{y})}{p_2(\bm{y})}\right] = \mathbb{E}_{A,B}\left[\log\frac{p(\bm{y}_{B}\given \bm{y}_{C_1})}{p(\bm{y}_{B}\given \bm{y}_{C_2})}\right] \\
&= \mathbb{E}_{B, C_1}\left[\log\frac{p(\bm{y}_{B}\given \bm{y}_{C_1})}{p(\bm{y}_{B}\given \bm{y}_{C_2})}\right] = \mathbb{E}_{C_1}\left\{\mathbb{E}_{B\given C_1}\left[\log\frac{p(\bm{y}_B\given \bm{y}_{C_1})}{p(\bm{y}_B\given \bm{y}_{C_2})}\right]\right\}
\geq 0\;.
\end{split}
\end{equation}
The equality $\mathbb{E}_{A,B}[\cdot] = \mathbb{E}_{B,C_1}[\cdot]$ in the last row follows from the fact that the argument is a function of diseases in $B$, $C_1$ and $C_2$ and, hence, in $B$ and $C_1$ because $C_2\subset C_1$. The argument given in \eqref{eq: KL_div_mult_cancer} is free of distributional assumptions and is linked to the submodularity of entropy and the ``information never hurts'' principle; see \cite{coverthomas91} and, more specifically, Eq.(18) in \cite{banerjee2020modeling}. Apart from providing a theoretical argument in favor of joint modeling, (\ref{eq: KL_div_mult_cancer}) also notes that models built upon hierarchical dependence structures depend upon the order in which the diseases enter the model. This motivates us to pursue model averaging over the different ordered models in a computationally efficient manner.

\subsection{Multivariate DAGAR Model}\label{sec: mdagar}
Modeling multiple diseases will introduce associations among the diseases and spatial dependence for each disease. Let $y_{ij}$ be a disease outcome of interest for disease $i$ in region $j$. For sake of clarity, we assume that $y_{ij}$ is a continuous variable (e.g., incidence rates) related to a set of explanatory variables through the regression model, 
\begin{equation}\label{eq: spatial_regression}
y_{ij} = \bm{x}_{ij}^\top\bm{\beta}_i + w_{ij} + e_{ij}\;,
\end{equation}
where $\bm{x}_{ij}$ is a $p_i\times 1$ vector of explanatory variables specific to disease $i$ within region $j$, $\bm{\beta}_i$ are the slopes corresponding to disease $i$, $w_{ij}$ is a random effect for disease $i$ in region $j$, and $e_{ij}\stackrel{ind}{\sim} N(0, (\sigma^2_i)^{-1})$ is the random noise arising from uncontrolled imperfections in the data. 

Part of the residual from the explanatory variables is captured by the spatial-temporal effect $w_{ij}$. Let $\bm{w}_i = (w_{i1}, w_{i2}, \dots, w_{ik})^\top$ for $i = 1,2,\ldots,q$. We adopt a hierarchical approach \citep[see, e.g.,][]{jin2005generalized}, where we specify the joint distribution of $\bm{w} = (\bm{w}_1^\top, \bm{w}_2^\top, \dots, \bm{w}_q^\top)^\top$ as $p(\bm{w}) = p(\bm{w}_1)\prod_{i=2}^q p(\bm{w}_i\given \bm{w}_{<i})$. We model $p(\bm{w}_1)$ and each of the conditional densities $p(\bm{w}_i\given \bm{w}_{<i})$ with $\bm{w}_{<i} = (\bm{w}_1^\top, \dots, \bm{w}_{i-1}^\top)^\top$ for $i \geq 2$ as univariate spatial models. The merits of this approach include simplicity and computational efficiency while ensuring that richness in structure is accommodated through the $p(\bm{w}_i\given \bm{w}_{<i})$'s. %One drawback, however, is that the model is specified through the order of appearance of the $\bm{w}_i$'s in the hierarchical model.     

We point out two important distinctions from \cite{jin2005generalized}: (i) instead of using conditional autoregression or CAR for the spatial dependence, we use DAGAR; and (ii) we apply a computationally efficient bridge sampling algorithms \citet{gronau2017tutorial} to compute the marginal posterior probabilities for each ordered model. The first distinction allows better interpretation of spatial autocorrelation than the CAR models. The second distinction is of immense practical value and makes this approach feasible for a much larger number of outcomes. Without this distinction, analysts would be dealing with $q!$ models for $q$ diseases and choose among them based upon a model-selection metric. That would be overly burdensome for more than 2 or 3 diseases. Details follow.

\subsection{A conditional multivariate DAGAR (MDAGAR) model}\label{sec: CMDAGAR}
The multivariate DAGAR (or MDAGAR) model is constructed as
\begin{align}
\bm{w}_1 = \bm{\epsilon}_1; \quad \bm{w}_i = \bm{A}_{i1}\bm{w}_1 + \bm{A}_{i2}\bm{w}_2 + \dots + \bm{A}_{i,i-1}\bm{w}_{i-1} + \bm{\epsilon}_i\; \mbox{ for }\; i=2,3,\ldots,q\;, \label{eq:w_q}
%\bm{w}_2 &= \bm{A}_{21}\bm{w}_1 + \bm{\epsilon}_2,  \hspace{1cm} \bm{\epsilon}_2 \sim N(\bm{0}, \tau_2^{-1}\bm{D}_2)\label{eq:w2} \\
%\vdots\\
\end{align}
where $\bm{\epsilon}_i \sim N(\bm{0}, \tau_i\bm{Q}(\rho_i))$ and $\tau_i \bm{Q}(\rho_i)$ are univariate DAGAR precision matrices with $\bm{B}$ and $\bm{F}$ as in \eqref{eq: B_and_F}. In \eqref{eq:w_q} we model $\bm{w}_1$ as a univariate DAGAR and, progressively, the conditional density of each $\bm{w}_i$ given $\bm{w}_1, \dots, \bm{w}_{i-1}$ is also as a DAGAR for $i=2,3,\ldots,q$. 

Each disease has its own distribution with its own spatial autocorrelation parameter. There are $q$ spatial autocorrelation parameters, $\{\rho_1, \rho_2, \dots, \rho_q\}$, corresponding to the $q$ diseases. Given the differences in the geographic variation of different diseases, this flexibility is desirable. %What is perhaps even more appealing is the ease of modeling associations among diseases while also modeling the spatial cross-covariances, i.e., the covariance between two diseases at two different, but neighboring, regions. 
Each matrix $\bm{A}_{ii'}$ in (\ref{eq:w_q}) with $i' = 1, \dots, i-1$ models the association between diseases $i$ and $i'$. We specify $\bm{A}_{ii'} = \eta_{0ii'}\bm{I}_k + \eta_{1ii'}\bm{M}$, where $\bm{M}$ is the binary adjacency matrix for the map, i.e., $m_{jj'}=1$ if $j'\sim j$ and $0$ otherwise. Coefficients $\eta_{0ii'}$ and $\eta_{1ii'}$ associate $w_{ij}$ with $w_{i'j}$ and $w_{i'j'}$. In other words, $\eta_{0ii'}$ is the diagonal element in $\bm{A}_{ii'}$, while $\eta_{1ii'}$ is the element in the $j$-th row and $j'$th column if $j' \sim j$. Therefore, for the joint distribution of $\bm{w}$, if $\bm{A}$ is the $kq\times kq$ strictly block-lower triangular matrix with $(ii')$-th block being $\bm{A}_{ii'} = \bm{O}$ whenever $i'\geq i$ and $\bm{\epsilon} = (\bm{\epsilon}_1^\top, \dots, \bm{\epsilon}_q^\top)^\top$, then (\ref{eq:w_q}) renders $\bm{w} = \bm{A}\bm{w} + \bm{\epsilon}$. 

Since $\bm{I} - \bm{A}$ is still lower triangular with $1$s on the diagonal, it is non-singular with $\det(\bm{I}-\bm{A})=1$. Writing $\bm{w} = \bm{(I - A)}^{-1}\bm{\epsilon}$, where $\bm{\epsilon} \sim N(\bm{0}, \bm{\Lambda})$ and the block diagonal matrix $\bm{\Lambda}$ has $\tau_1\bm{Q}(\rho_1), \dots, \tau_q\bm{Q}(\rho_q)$ on the diagonal, we obtain $\bm{w} \sim N(\bm{0}, \bm{Q}_w)$ for $\bm{\rho} = (\rho_1, \dots, \rho_q)^\top$ with
\begin{align}\label{eq:cov}
\bm{Q}_w = \bm{(I - A)}^\top \bm{\Lambda}\bm{(I - A)}\;.
\end{align}
We say that $\bm{w}$ follows MDAGAR if $\bm{w}\sim N(\bzero, \bm{Q}_w)$.
%and the covariance matrix $\bm{Q}^{-1}(\bm{\rho})$ is $\bm{(I - A_q)}^{-1}\mbox{Cov}(\bm{\epsilon})\bm{(I - A_q)}^{-\top} = \bm{(I -  A_q)}^{-1}\bm{\Lambda_q}\bm{(I - A_q)}^{-\top}$.
\begin{comment}
In particular, with $q=2$, the joint distribution of $\bm{w} = (\bm{w}_1^{\top}, \bm{w}_2^{\top})^{\top}$ is $p(\bm{w}_1,\bm{w}_2) = N(\bm{w}_1\given 0, \tau_1 \bm{Q}(\rho_1)) \times N(\bm{w}_2\given \bm{A}_{21}\bm{w}_1, \tau_{2} \bm{Q}(\rho_2))$,
%	\begin{equation}\label{eq: bivariate_DAGAR}
%		p(\bm{w}_1,\bm{w}_2) = N(\bm{w}_1\given 0, \tau_1 \bm{Q}(\rho_1)) \times %N(\bm{w}_2\given \bm{A}_{21}\bm{w}_1, \tau_{2} \bm{Q}(\rho_2))\;, 
%	\end{equation}
where $\bm{w}$ has the precision matrix
\begin{equation}\label{eq: bivariate_DAGAR_precision}
\bm{Q}_w = \begin{bmatrix} 
\tau_1 \bm{Q}(\rho_1) + \tau_2 \bm{A}_{21}^{\top}\bm{Q}(\rho_2)\bm{A}_{21} & \tau_2 \bm{A}_{21}^{\top} \bm{Q}(\rho_2) \\ 
\tau_2 \bm{Q}(\rho_2)\bm{A}_{21} & \tau_2 \bm{Q}(\rho_2) \end{bmatrix}                                
\end{equation}
\end{comment}
Interpretation of $\rho_1, \ldots, \rho_q$ is clear: $\rho_1$ measures the spatial association for the first disease, while $\rho_i$, $i \geq 2$, is the residual spatial correlation in the disease $i$ after accounting for the first $i-1$ diseases. Similarly, $\tau_1$ is the spatial precision for the first disease, while $\tau_i$, $i \geq 2$, is the residual spatial precision for disease $i$ after accounting for the first $i-1$ diseases.

\subsubsection{Model Implementation} \label{Implementation}
We extend (\ref{eq: spatial_regression}) to the following Bayesian hierarchical framework with the posterior distribution $p(\bm{\beta}, \bm{w}, \bm{\eta}, \bm{\rho}, \bm{\tau}, \bm{\sigma} \given \bm{y})$ proportional to
\begin{align}\label{eq: bivariate_DAGAR_bhm}
& p(\bm{\rho}) \times p(\bm{\eta}) \times \prod_{i=1}^q \left\{IG(1/\tau_i\given a_{\tau}, b_{\tau}) \times IG(\sigma_i^2 \given a_{\sigma}, b_{\sigma}) \times N(\bm{\beta}_i\given \bm{\mu}_{\beta}, \bm{V}_{\beta}^{-1})\right\} \nonumber\\ 
&\qquad \times N(\bm{w}\given \bm{0}, \bm{Q}_w) \times \prod_{i=1}^q\prod_{j=1}^k N(y_{ij}\given \bm{x}_{ij}^{\top}\bm{\beta}_i + w_{ij},1/\sigma_i^2)\;,
\end{align} 
where $\bm{\beta} = (\bm{\beta}_1^\top, \bm{\beta}_2^\top, \dots, \bm{\beta}_q^\top)^\top$, $\bm{\tau}=\{\tau_1,\tau_2,\dots,\tau_q\}$, $\bm{\sigma} = \{\sigma^2_1, \sigma^2_2, \dots, \sigma^2_q\}$ and $\bm{\eta} = \{\bm{\eta}_2,\bm{\eta}_3, \dots,\\ \bm{\eta}_q\}$ with $\bm{\eta}_{i} = (\bm{\eta}_{i1}^\top,\bm{\eta}_{i2}^\top, \dots, \bm{\eta}_{i,i-1}^\top)^\top$ and $\bm{\eta}_{ii'} = (\eta_{0ii'},\eta_{1ii'})^\top$ for $i = 2, \dots, q$ and $i' = 1, \dots, i-1$. For variance parameters $1/\tau_i$ and $\sigma_i^2$,  $IG(\cdot\given a,b)$ is the inverse-gamma distribution with shape and rate parameters $a$ and $b$, respectively. For each element in $\bm{\eta}_i$ we choose a normal prior $N(\mu_{ij}, \sigma_{\eta_{ij}}^2)$, while the prior $N(\bm{w}\given \bm{0},\bm{Q}_w)$ %derived from \eqref{eq:w_q} 
can also be written as %$p(\bm{w}_1 | \tau_1, \rho_1) \times p(\bm{w}_2 | \bm{w}_1, \bm{\eta}_{2} , \tau_2, \rho_2) \times \dots \times p(\bm{w}_q | \bm{w}_1,\dots, \bm{w}_{q-1}, \bm{\eta}_{q} , \tau_q, \rho_q)$, i.e.
\begin{multline}
p(\bm{w} |\bm{\tau}, \bm{\eta}_2,\dots, \bm{\eta}_q,  \bm{\rho}) \propto
\tau_1^{\frac{k}{2}}|\bm{Q}(\rho_1)|^{\frac{1}{2}}\exp{\left\{-\frac{\tau_1}{2}\bm{w}_1^\top\bm{Q}(\rho_1)\bm{w}_1\right\}}\\
\times \prod_{i=2}^q\tau_i^{\frac{k}{2}}|\bm{Q}(\rho_i)|^{\frac{1}{2}}\exp{\left\{-\frac{\tau_i}{2}(\bm{w}_i-\sum_{i'=1}^{i-1}\bm{A}_{ii'}\bm{w}_{i'})^\top \bm{Q}(\rho_i)(\bm{w}_i-\sum_{i'=1}^{i-1}\bm{A}_{ii'}\bm{w}_{i'})\right\}}\;,
\end{multline}
where $\det(\bm{Q}(\rho_i)) = \prod_{j=1}^k \lambda_{ij}$, and $\bm{w}_i^T\bm{Q}(\rho_i)\bm{w}_i = \lambda_{i1}w_{i1}^2 + \sum_{j=2}^k\lambda_{ij}(w_{ij} - \sum_{j' \in N(j)}b_{ijj'}w_{ij'})^2$.

We sample the parameters from the posterior distribution in \eqref{eq: bivariate_DAGAR_bhm} using Markov chain Monte Carlo (MCMC) with Gibbs sampling and random walk metropolis \citep{gamerman2006markov} as implemented in the \texttt{rjags} package within the \texttt{R} statistical computing environment. %The parameters $\bm{w}, \bm{\beta}, \bm{\sigma}, \bm{\tau}, \bm{\eta}_2,\dots, \bm{\eta}_q$ are updated through a Gibbs sampler using their full conditional distributions, while $\bm{\rho}$ are updated as a single block using a Metropolis random walk. 
%Since $\bm{\rho}$ is embedded in the joint precision matrix $\bm{Q}_w$, the closed form of the full conditional distribution does not exist. Hence, at each Gibbs iteration, we generate one sample of $\bm{\rho}$ randomly using standard normal for each $\rho_i$ and decide whether to accept it depending on $N(\bm{w} | \bm{0}, \bm{Q}_w) \times p(\bm{\rho})$ given current values of $\bm{w}, \bm{\tau}, \bm{\eta}_2,\dots, \bm{\eta}_q$ in a Metropolis step. 
Section~\ref{sup_implementation} presents details on the the MCMC updating scheme. %Overall, the MCMC chains are designed to run for 30000 iterations with the first 15000 iterations as ``burn in" and the last 15000 iterations as MCMC samples are used for posterior summarization. The computation is only coded up manually for MDAGAR model in simulation, otherwise models are implemented in the \texttt{rjags} package within the \texttt{R} statistical computing environment

\subsection{Model Selection via Bridge Sampling}\label{selection}
It is clear from \eqref{eq:w_q} that each ordering of diseases in MDAGAR will produce a different model. For the bivariate situation, it is convenient to compare only two models (orders) by the significance of parameter estimates as well as model performance. However, when there are more than two diseases involved in the model, at least six models (for three diseases) will be fitted and comparing all models become cumbersome or even impracticable. 

Instead, we pursue model averaging of MDAGAR models. Given a set of $T=q!$ candidate models, say $M_1, \ldots, M_T$, Bayesian model selection and model averaging calculates 
\begin{align}
p(M = M_t|\bm{y}) = \frac{p(\bm{y}|M = M_t)p(M = M_t)}{\sum_{j=1}^Tp(\bm{y}|M=M_j)p(M=M_j)},\label{eq:bma}
\end{align}
for $t = 1, \dots, T$ \citep{hoeting1999bayesian}. Computing the marginal likelihood $p(\bm{y}\given M_t)$ in \eqref{eq:bma} is challenging. Methods such as importance sampling \citep{perrakis2014use} and generalized harmonic mean \citep{gelfand1994bayesian} have been proposed as stable estimators with finite variance, but finding the required importance density with strong constraints on the tail behavior relative to the posterior distribution is often challenging. Bridge sampling estimates the marginal likelihood (i.e. the normalizing constant) by combining samples from two distributions: a bridge function $h(\cdot)$ and a proposal distribution $g(\cdot)$ \citep{gronau2017bridgesampling}. Let $\bm{\theta}_t = \left\{\bm{\beta}_t, \bm{\sigma}_t, \bm{\rho}_t, \bm{\tau}_t, \bm{\eta}_{2,t},\dots, \bm{\eta}_{q,t}\right\}$ be the set of parameters in model $M_t$ with prior $p(\bm{\theta}_t \given M_t)$ as defined in the first row of \eqref{eq: bivariate_DAGAR_bhm}. Based on the identity,
\begin{align*}
1 = \frac{\int p(\bm{y} | \bm{\theta}_t, M_t)p(\bm{\theta}_t|M_t)h(\bm{\theta}_t|M_t)g(\bm{\theta}_t|M_t)d\bm{\theta}_t}{\int p(\bm{y} | \bm{\theta}_t, M_t)p(\bm{\theta}_t|M_t)h(\bm{\theta}_t|M_t)g(\bm{\theta}_t|M_t)d\bm{\theta}_t},
\end{align*}
a current version of the bridge sampling estimator is
\begin{equation}\label{eq:est}
\begin{split}
p(\bm{y}|M = M_t) &= \frac{E_{g(\bm{\theta}_t|M_t)}[p(\bm{y} | \bm{\theta}_t, M_t)p(\bm{\theta}_t|M_t)h(\bm{\theta}_t|M_t)]}{E_{p(\bm{\theta}_t|\bm{y}, M_t)}[h(\bm{\theta}_t|M_t)g(\bm{\theta}_t|M_t)]}\\
&\approx \frac{\frac{1}{N_2}\sum_{i=1}^{N_2}p(\bm{y} | \tilde{\bm{\theta}}_{t,i}, M_t)p(\tilde{\bm{\theta}}_{t,i}|M_t)h(\tilde{\bm{\theta}}_{t,i}|M_t)}{\frac{1}{N_1}\sum_{j=1}^{N_1}h(\bm{\theta}_{t,j}^\star|M_t)g(\bm{\theta}_{t,j}^\star|M_t)}
\end{split}
\end{equation}
where $\bm{\theta}_{t,j}^\star \sim p(\bm{\theta}_t\given\bm{y}, M_t), j = 1, \dots, N_1$, are $N_1$ posterior samples and $\tilde{\bm{\theta}}_{t,i} \sim g(\bm{\theta}_t|M_t), i = 1, \dots, N_2$, are $N_2$ samples drawn from the proposal distribution \citep{gronau2017tutorial}. The likelihood $p(\bm{y} \given \bm{\theta}_t, M = M_t)$ is obtained by integrating out $\bm{w}$ from \eqref{eq: bivariate_DAGAR_bhm} as
\begin{align} 
N(\bm{y} \given \bm{X\beta}, \big[\bm{Q}_{w}^{-1}(\bm{\rho}_t, \bm{\tau}_t, \bm{\eta}_{2,t}, \dots, \bm{\eta}_{q,t}) + diag(\bm{\sigma}_t)\otimes \bm{I}_k\big]^{-1}), \label{eq:marginal_likelihood}
\end{align} 
given that $\bm{y} = (\bm{y}_1^\top, \dots, \bm{y}_q^\top)^\top$ with $\bm{y}_i = (y_{i1}, y_{i2}, \dots, y_{ik})^\top$, $diag(\bm{\sigma})$ is a diagonal matrix with $\sigma_i^2, i =1,\dots,q$, on the diagonal, and $\bm{X}$ is the design matrix with $\bm{X}_i$ as block diagonal where $\bm{X}_i = (\bm{x}_{i1}, \bm{x}_{i2}, \dots, \bm{x}_{ik})^\top$. The bridge function $h(\bm{\theta}_t|M_t)$ is specified by the optimal choice proposed in \citet{meng1996simulating},
\begin{align}
h(\bm{\theta}_t|M_t) = C \frac{1}{s_1p(\bm{y} | \bm{\theta}_t, M_t)p(\bm{\theta}_t|M_t) + s_2p(\bm{y}|M_t)g(\bm{\theta}_t|M_t)}\label{eq:bridge_fun}
\end{align}
where $C$ is a constant. Inserting \eqref{eq:bridge_fun} in \eqref{eq:est} yields the estimate of $p(\bm{y}|M=M_t)$ after convergence of an iterative scheme \citep{meng1996simulating} as
\begin{align}
\hat{p}(\bm{y}|M_t)^{(t+1)} = \frac{\frac{1}{N_2}\sum_{i=1}^{N_2}\frac{l_{2,i}}{s_1l_{2,i} + s_2\hat{p}(\bm{y}|M_t)^{(t)}}}{\frac{1}{N_1}\sum_{j=1}^{N_1}\frac{1}{s_1l_{1,j} + s_2\hat{p}(\bm{y}|M_t)^{(t)}}}
\end{align}
where $l_{1,j} = \frac{p(\bm{y} | \bm{\theta}_{t,j}^\star, M_t)p(\bm{\theta}_{t,j}^\star|M_t)}{g(\bm{\theta}_{t,j}^\star|M_t)}$, $l_{2,i} = \frac{p(\bm{y} | \tilde{\bm{\theta}}_{t,i}, M_t)p(\tilde{\bm{\theta}}_{t,i}|M_t)}{g(\tilde{\bm{\theta}}_{t,i}|M_t)}$, $s_1 = \frac{N_1}{N_1+N_2}$ and $s_2 = \frac{N_2}{N_1+N_2}$. %Using \texttt{JAGS} output from \ref{Implementation} as MCMC posterior samples, this algorithm is implemented in an \texttt{R} package \texttt{bridgesampling} with two options for the proposal function $g(\bm{\theta}_t|M_t)$, including a multivariate normal distribution as default and a standard multivariate normal distribution in combination with a warped posterior distribution \citep{gronau2017bridgesampling}.

Given the log marginal likelihood estimates from \texttt{bridgesampling}, the posterior model probability for each model is calculated from \eqref{eq:bma} by setting prior probability of each model $p(M=M_t)$. For Bayesian model averaging (BMA), the model averaged posterior distribution of a quantity of interest $\Delta$ is obtained as $p(\Delta \given \bm{y}) = \sum_{t=1}^T p(\Delta \given M = M_t, \bm{y}) p(M = M_t \given \bm{y})$
\citep{hoeting1999bayesian}, and the posterior mean is 
\begin{align}\label{eq: ema}
E(\Delta \given \bm{y}) = \sum_{t=1}^T E(\Delta \given M = M_t, \bm{y}) p(M = M_t \given \bm{y})\;.
\end{align}
Setting $\Delta = \{\bm{\beta, w}\}$ fetches us the model averaged posterior estimates for spatial random effects as well as calculating the posterior mean incidence rates as discussed in Section~\ref{data}.

\section{Simulation}\label{simulation}
We simulate two different experiments. The first experiment is designed to evaluate MDAGAR's inferential performance against GMCAR. The second experiment aims to ascertain the effectiveness of the bridge sampling algorithm (\ref{selection}) in preferring models with a correct ``ordering'' of the diseases in the model.  

\subsection{Data generation}\label{gen}
We compare MDAGAR's inferential performance with GMCAR \citep{jin2005generalized}. We choose the 48 states of the contiguous United States as our underlying map, where two states are treated as neighbors if they share a common geographic boundary. We generated our outcomes $y_{ij}$ using the model in \eqref{eq: spatial_regression} with $q=2$, i.e., two outcomes, and two covariates, $\bm{x}_{1j}$ and $\bm{x}_{2j}$, with $p_1=2$ and $p_2=3$. We fixed the values of the covariates after generating them from $N(\bm{0}, \bm{I}_{p_i})$, $i=1,2$, independent across regions. The regression slopes were set to $\bm{\beta}_1 = (1, 5)^{\top}$ and $\bm{\beta}_2 = (2,4,5)^{\top}$. %and we specify $\sigma_{1}^2 = \sigma_{2}^2 = 0.4$ in the error distribution in (\ref{eq: spatial_regression}).  

Turning to the spatial random effects, we generated values of $\bm{w}= \left(\bm{w}_1^{\top}, \bm{w}_2^{\top}\right)^{\top}$ from a $N(\bm{0},\bm{Q}_w)$ distribution, where the precision matrix is
\begin{equation}\label{eq: bivariate_DAGAR_precision}
\bm{Q}_w = 
\begin{bmatrix} 
\tau_1 \bm{Q}(\rho_1) + \tau_2 \bm{A}_{21}^{\top}\bm{Q}(\rho_2)\bm{A}_{21} & \tau_2 \bm{A}_{21}^{\top} \bm{Q}(\rho_2) \\ 
\tau_2 \bm{Q}(\rho_2)\bm{A}_{21} & \tau_2 \bm{Q}(\rho_2) 
\end{bmatrix}\;.                                
\end{equation}
We set $\tau_1 = \tau_2 = 0.25$, $\rho_1 = 0.2$ and $\rho_2 = 0.8$ in (\ref{eq: bivariate_DAGAR_precision}) and take $\bm{Q}(\rho_i) = \bm{D}(\rho_i)^{-1}$, where $\bm{D}(\rho_i) = \exp(-\phi_i d(j, j'))$, $\phi_i = -\log(\rho_i)$ is the spatial decay for disease $i$ and $d(j, j')$ refers to the distance between the embedding of the $j$th and $j'$th vertex. The vertices are embedded on the Euclidean plane and the centroid of each state is used to create the distance matrix. Using this exponential covariance matrix to generate the data offers a ``neutral'' ground to compare the performance of MDAGAR with GMCAR. We specified $\bm{A}_{12}$ using fixed values of $\bm{\eta} = \{\eta_{021},\eta_{121}\}$. Here, we considered three sets of values for $\bm{\eta}$ to correspond to low, medium and high correlation among diseases. We fixed $\bm{\eta} = \{0.05, 0.1\}$ to ensure an average correlation of 0.15 (range 0.072 - 0.31); $\bm{\eta} = \{0.5,0.3\}$ with an average correlation of 0.55 (range 0.45 - 0.74); and $\bm{\eta} = \{2.5,0.5\}$ with a mean correlation of 0.89 (range 0.84 - 0.94). We generated $w_{ij}$'s for each of the above specifications for $\bm{\eta}$ and, with the values of $w_{ij}$ generated as above, we generated the outcome $y_{ij}\sim N(\bm{x}_{ij}^{\top}\bbeta_i + w_{ij}, 1/\sigma_i^2)$, where $\sigma_{1}^2 = \sigma_{2}^2 = 0.4$. We repeat the above procedure to replicate $85$ data sets for each of the three specifications of $\bm{\eta}$. 

For our second experiment we generate a data set with $q=3$ cancers. We extend the above setup to include one more disease. We generate $y_{ij}$'s from \eqref{eq: spatial_regression} with the value of $\bm{x}_{3j}$ fixed after being generated from $N(\bm{0}, \bm{I}_3)$, $\bm{\beta}_3 = (5, 3, 6)^{\top}$ and  $\sigma_{3}^2 = 0.4$. 
%We will generate $\bm{w} = (\bm{w}_1^{\top}, \bm{w}_2^{\top}, \bm{w}_3^{\top})$ using a directed acyclic graph (DAG) that specifies the order. 
Let $[i,j,k]$ denote the model $p(\bm{w}_i) \times p(\bm{w}_j \given \bm{w}_i) \times p(\bm{w}_k \given \bm{w}_j, \bm{w}_i)$. For three diseases the six resulting models are denoted as $M_1 = [1,2,3]$, $M_2 = [1,3,2]$, $M_3 = [2,1,3]$, $M_4 = [2,3,1]$, $M_5 = [3,1,2]$ and $M_6 = [3,2,1]$.

Each of the six models imply a corresponding joint distribution $\bm{w}\sim N(\bm{0}, \bm{Q}_w)$ which is used to generate the $w_{ij}$'s. Let the parenthesized suffix $(i)$ denote the disease in the $i$th order. For example, in $M_2 = [1,3,2]$, we write $\bm{w}$ in the form of \eqref{eq:w_q} as
\begin{align*}
\bm{w}_1 \sim \epsilon_{(1)} ; \quad \bm{w}_3 = \bm{A}_{(21)}\bm{w}_1 + \epsilon_{(2)}; \quad \bm{w}_2 = \bm{A}_{(31)}\bm{w}_1 + \bm{A}_{(32)}\bm{w}_3 + \epsilon_{(3)}\;,
\end{align*}
where $\epsilon_{(i)} \sim N(\bm{0}, \tau_{(i)}\bm{Q}(\rho_{(i)}))$ with $\bm{Q}(\rho_{(i)}) = \bm{D}(\rho_{(i)})^{-1}$ as in the first experiment, and $\bm{A}_{(ii')} = \eta_{0(ii')} \bm{I}+ \eta_{1(ii')}\bm{M}$ is the coefficient matrix associating random effects for diseases in the $i$th and $i'$th order. We set $\tau_{(1)} = \tau_{(2)} = \tau_{(3)} = 0.25$, $\rho_{(1)} = 0.2$, $\rho_{(2)} = 0.8$, $\rho_{(3)} = 0.5$, $\eta_{0(21)} = 0.5$, $\eta_{1(21)} = 0.3$, $\eta_{0(31)} = 1$, $\eta_{1(31)} = 0.6$, $\eta_{0(32)} = 1.5$, and $\eta_{1(32)} = 0.9$ to completely specify $\bm{Q}_w$ for each of the 6 models. For each $M_i$, we generate $50$ datasets by first generating $\bm{w}\sim N(\bm{0}, \bm{Q}_w)$ and then generating $y_{ij}$'s from (\ref{eq: spatial_regression}) using the specifications described above.

\subsection{Comparisons between MDAGAR and GMCAR}\label{modelcom}
In our first experiment we analyzed the $85$ replicated datasets using (\ref{eq: bivariate_DAGAR_bhm}) with
\begin{equation}\label{eq: priors_sim_1}
%\begin{split}
%& 
p(\bm{\rho}) \times p(\bm{\eta}) \propto \prod_{i=1}^{q=2} \left\{Unif(\rho_i\given 0,1)\right\}
%&\qquad\qquad\qquad \qquad 
\times 
N(\bm{\eta}_{21}\given \bm{0}, 0.01\bm{I}_2)\;, 
% \end{split}
\end{equation}
where $\bm{\eta}_{21} = (\eta_{021}, \eta_{121})^{\top}$ and $Unif$ is the Uniform density. Prior specifications are completed by setting $a_\tau = 2$, $b_\tau = 8$, $a_{\sigma} = 2$, $b_{\sigma} = 0.4$, $\bm{\mu}_{\beta} = \bm{0}$, $\bm{V}_{\beta} = 1000\bm{I}$ in (\ref{eq: bivariate_DAGAR_bhm}). 
%For variance parameters $\sigma_i^2$ and $1/\tau_i$ are centered at their true values for ease of convergence.
Note that the same set of priors were used for both MDAGAR and GMCAR as they have the same number of parameters with similar interpretations.

\begin{figure}[h]
	\begin{subfigure}[t]{0.35\textwidth}
		\centering
		\includegraphics[scale=0.3]{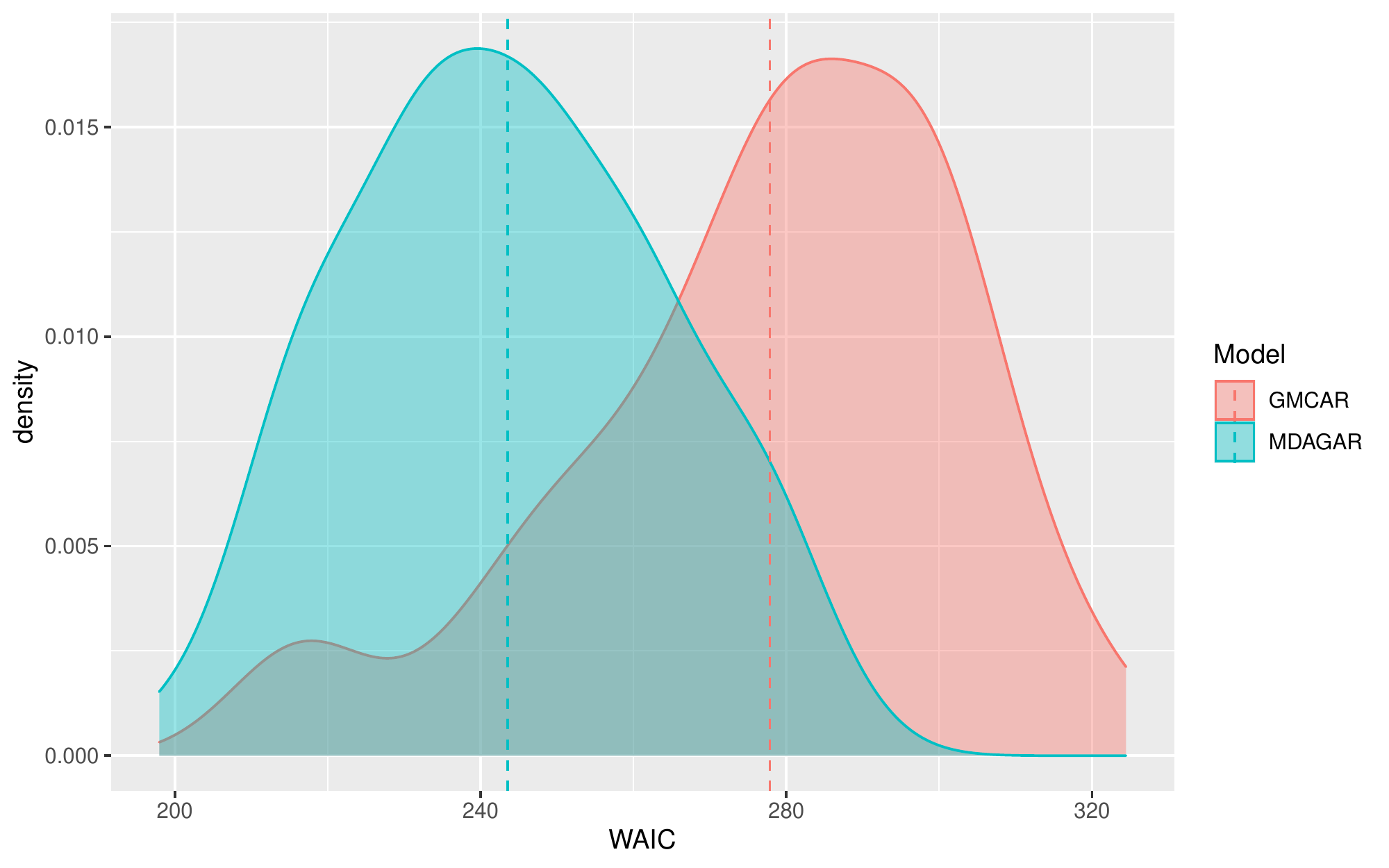}
		\caption{WAIC: low}\label{fig: waicl}
	\end{subfigure} 
	\hskip -0.8cm \begin{subfigure}[t]{0.35\textwidth}
		\centering
		\includegraphics[scale=0.3]{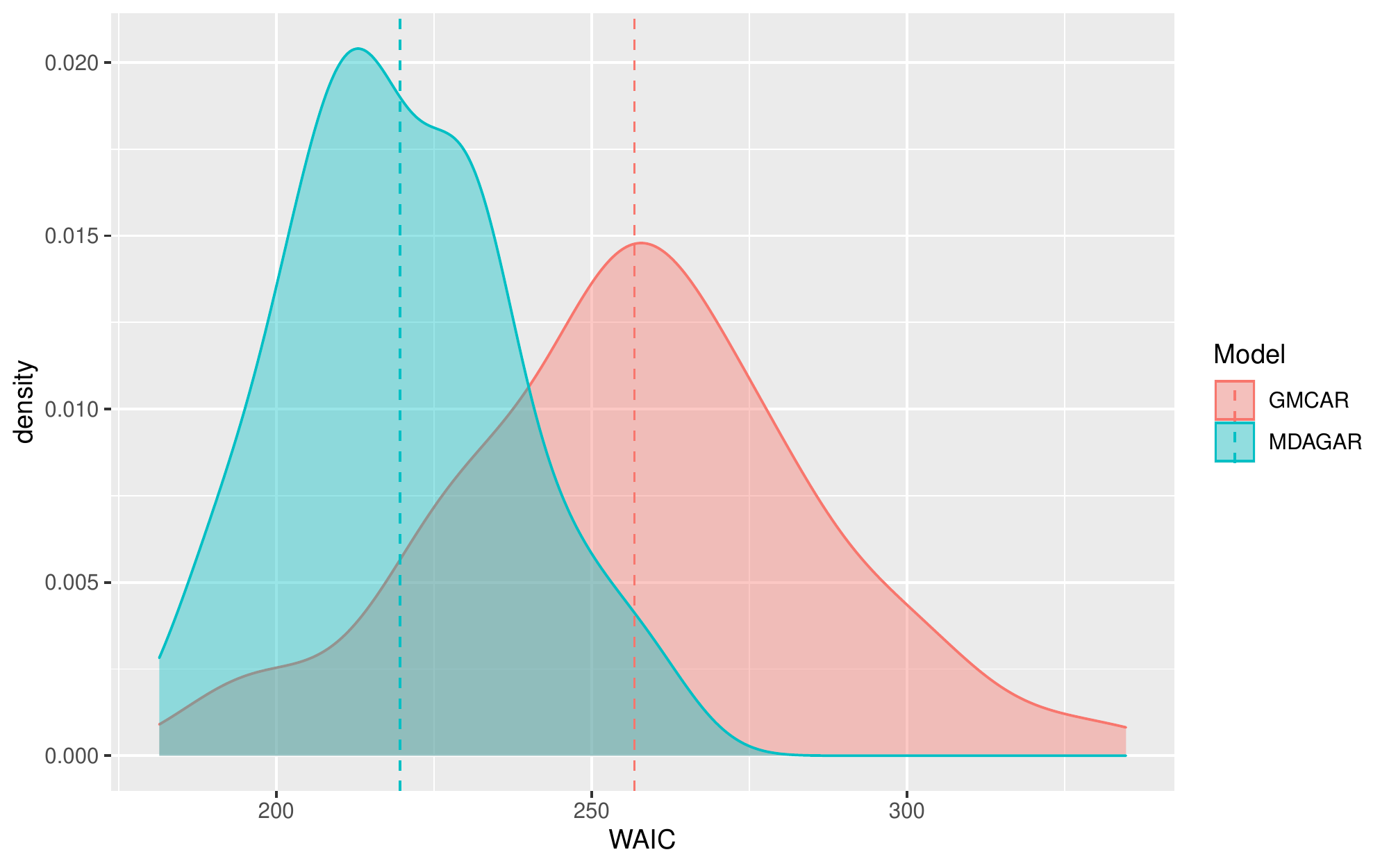}
		\caption{WAIC: medium}\label{fig: waicm}
	\end{subfigure} 
	\hskip -0.8cm \begin{subfigure}[t]{0.35\textwidth}
		\centering
		\includegraphics[scale=0.3]{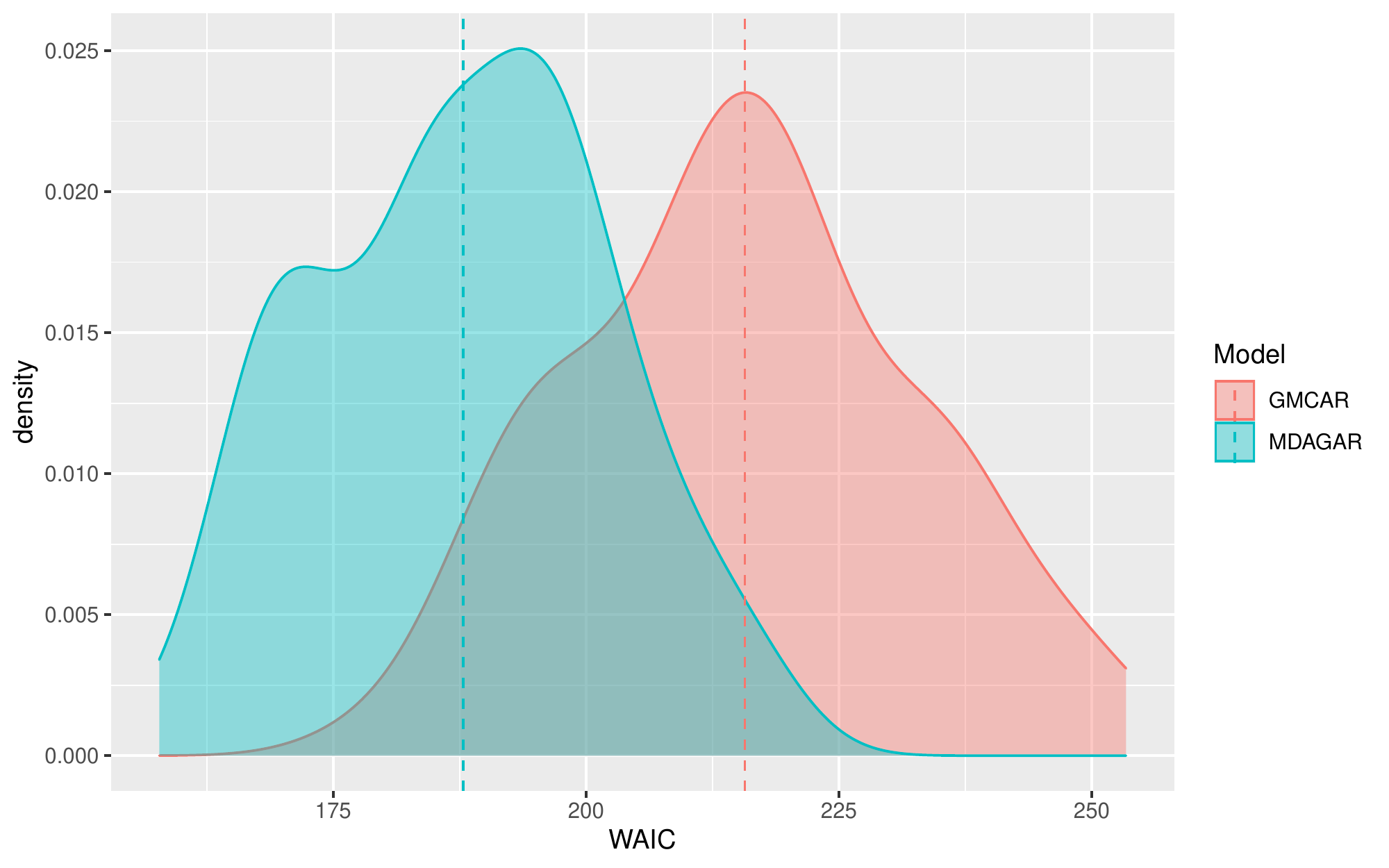}
		\caption{WAIC: high}\label{fig: waich}
	\end{subfigure}
	\vfill
	\begin{subfigure}[t]{0.35\textwidth}
		\centering
		\includegraphics[scale=0.3]{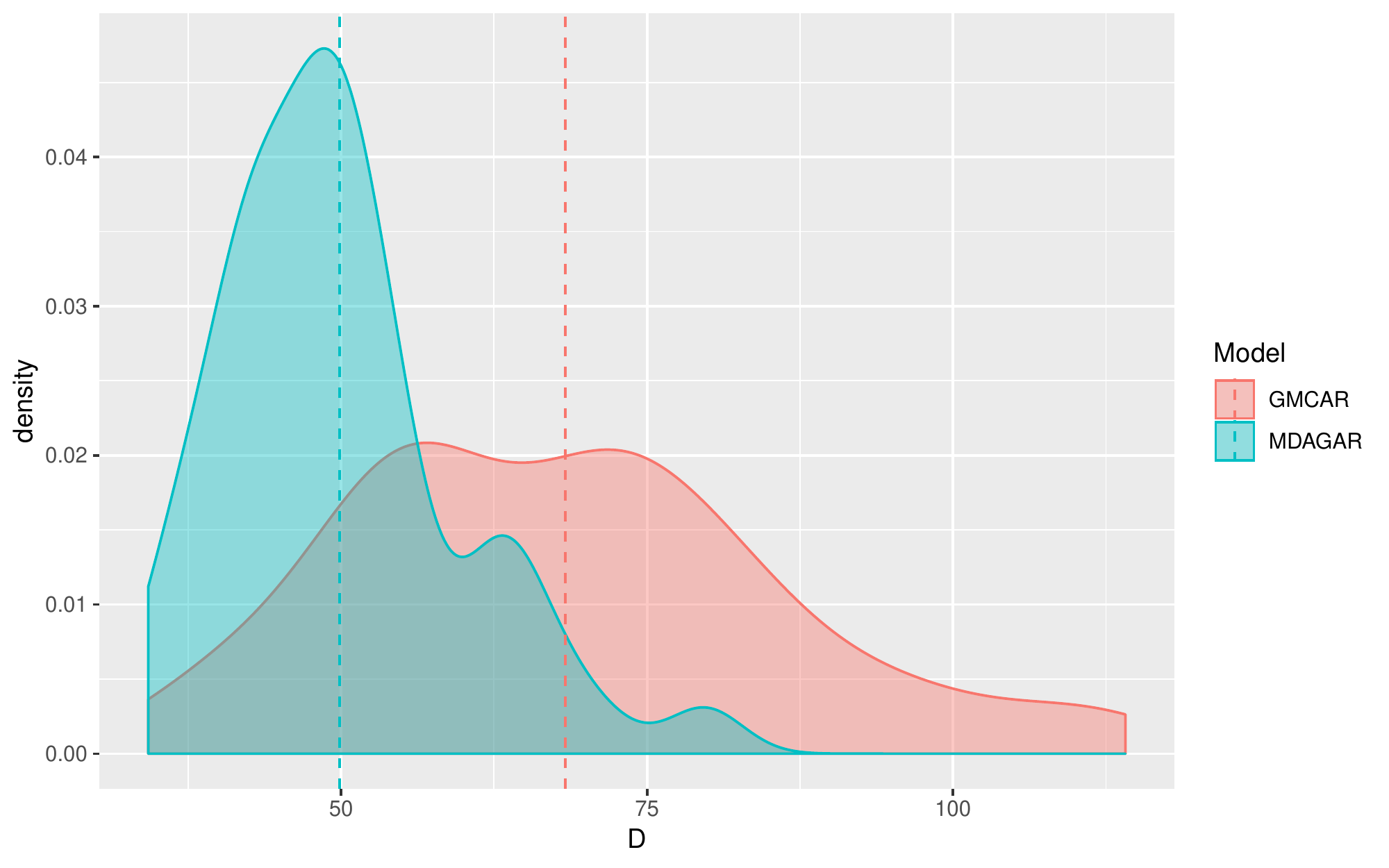}
		\caption{Score $D$: low}\label{fig: dl}
	\end{subfigure} 
	\hskip -0.8cm \begin{subfigure}[t]{0.35\textwidth}
		\centering
		\includegraphics[scale=0.3]{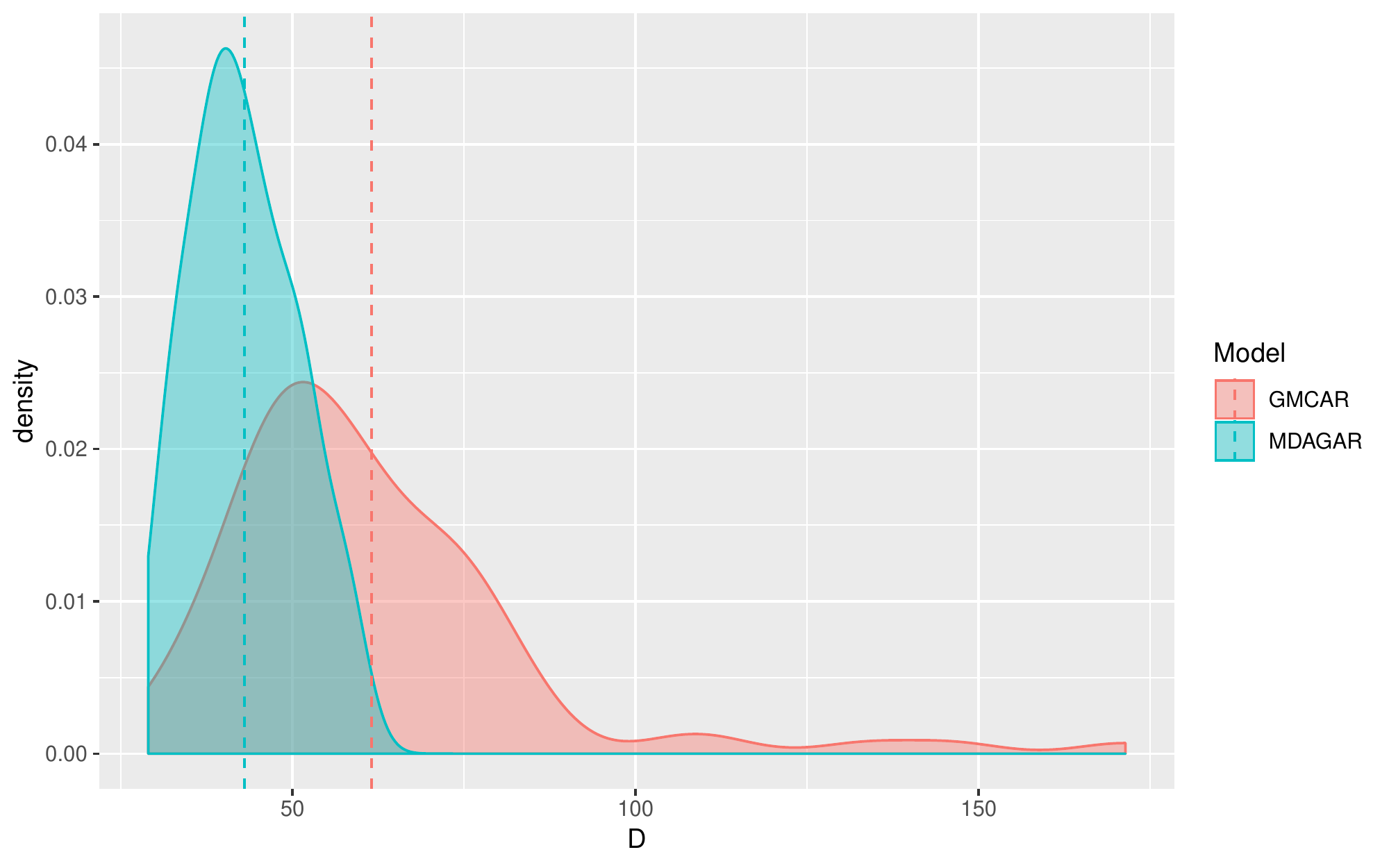}
		\caption{Score $D$: medium}\label{fig: dm}
	\end{subfigure} 
	\hskip -0.8cm \begin{subfigure}[t]{0.35\textwidth}
		\centering
		\includegraphics[scale=0.3]{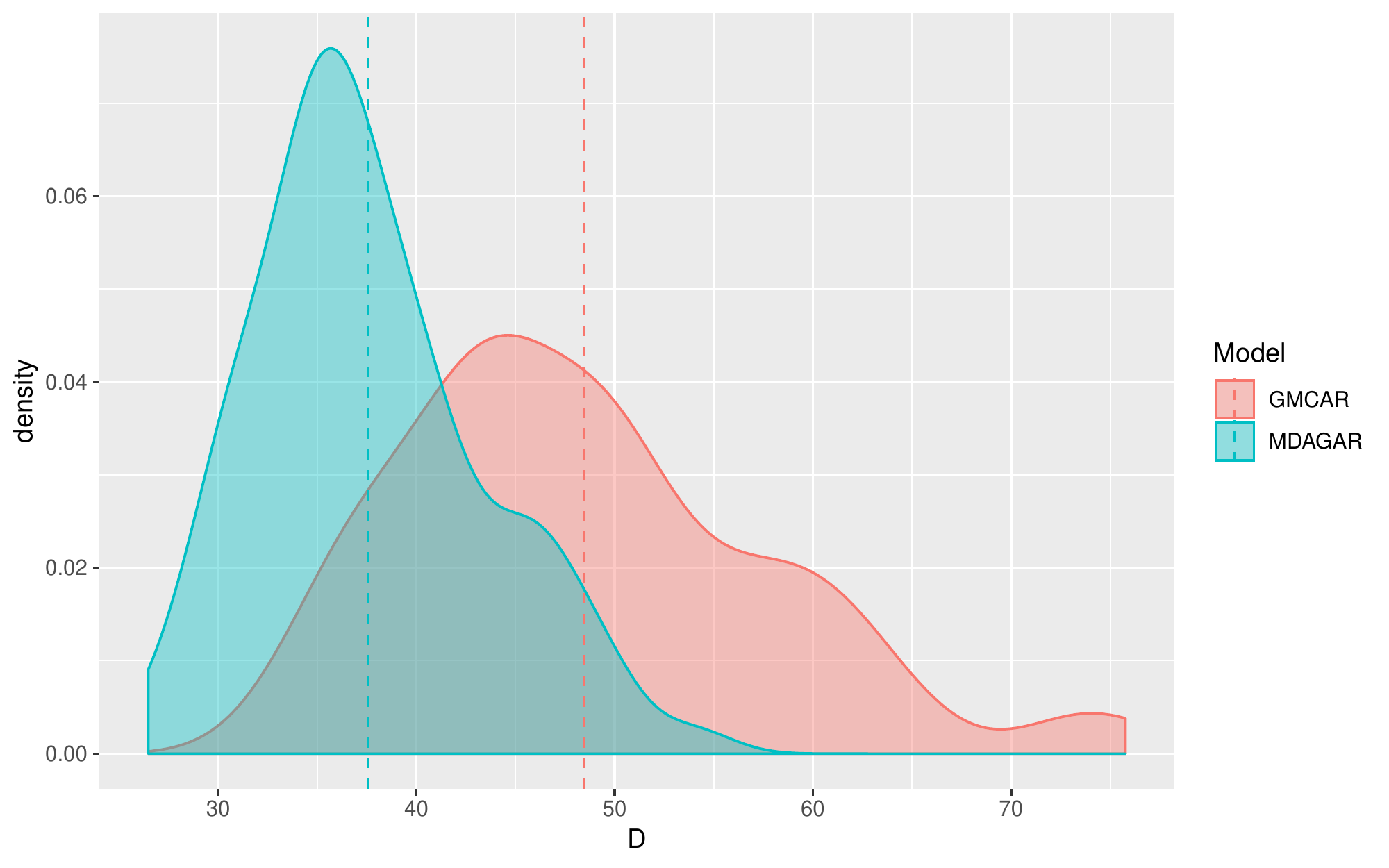}
		\caption{Score $D$: high}\label{fig: dh}
	\end{subfigure}
	\caption{Density plots for WAICs and $D$ scores over 85 datasets. Density plots of WAIC for MDAGAR (blue) and GMCAR (red) models with low, medium and high correlation are shown in (a), (b) and (c) respectively, while (d)--(f) are the corresponding density plots for $D$ scores. The dotted vertical line shows the mean for WAIC and D in each plot.}\label{fig: waic}
	%		\end{adjustwidth}
\end{figure}

We compare models using the Widely Applicable Information Criterion (WAIC) \citep{watanabe2010asymptotic, gelman2014understanding} and a model comparison score $D$ based on a balanced loss function for replicated data \citep{gelfand1998model}. Both WAIC and $D$ reward goodness of fit and penalize model complexity. Details on how these metrics are computed are provided in \ref{sup_sim}. In addition, we also computed the average mean squared error (AMSE) of the spatial random effects estimated from each of the $85$ data sets. We found the mean (standard deviation) of the AMSEs to be 1.69 (0.034) from the 85 low-correlation datasets, 1.47 (0.030) from the 85 medium-correlation datasets, and 2.35 (0.059) from the 85 high-correlation datasets. The corresponding numbers for GMCAR were 1.83 (0.033), 1.59 (0.031), and 2.14 (0.050), respectively. 
%Table~\ref{tab: mse} presents the mean, the $2.5$th and $97.5$th quantiles of the AMSEs and WAICs over the 85 data sets when the correlation between the two diseases is low, medium and high. 
The MDAGAR tends to have smaller AMSE for low and medium correlations, while GMCAR has lower AMSE when the correlations are high, although the differences are not significant. %MDAGAR excels over GMCAR in terms of WAIC scores in all three scenarios. 
\begin{comment}
\begin{table}[h]
\centering
\caption{Comparison of average mean squared error (AMSE) with associated Monte Carlo standard errors (SE) for random effects and WAIC with corresponding $95\%$ credible interval (CI) for MDAGAR and GMCAR models}\label{tab: mse}
\begin{tabular}{lccc}
\hline
& Low   & Medium & High \\
\hline
AMSE (SE) &       &       &  \\
MDAGAR & 1.69 (0.034) & 1.47 (0.030) & 2.35 (0.059) \\
GMCAR  & 1.83 (0.033) & 1.59 (0.031) & 2.14 (0.050) \\
\hline
WAIC (CI) &       &       &  \\
MDAGAR & 243.5 (212.1, 279.6) & 219.6 (188.4, 257.0) & 187.9 (165.2, 214.6) \\
GMCAR  & 277.8 (217.5, 315.8) & 256.8 (198.2, 313.2) & 215.7 (187.2, 247.3) \\
\hline
\end{tabular}
\end{table}
\end{comment}
We also compute the WAICs and $D$ scores for each simulated data set. Figure~\ref{fig: waic} plots the values of WAICs ((\subref{fig: waicl})--(\subref{fig: waich})) and $D$ scores ((\subref{fig: dl})--(\subref{fig: dh})) for the $85$ data sets corresponding to each of the three correlation settings. Here, MDAGAR outperforms GMCAR in all three correlation settings with respect to both WAICs and $D$ scores.

\begin{figure}[h]
	\begin{subfigure}[t]{0.3\textwidth}
		\centering
		\includegraphics[scale=0.25]{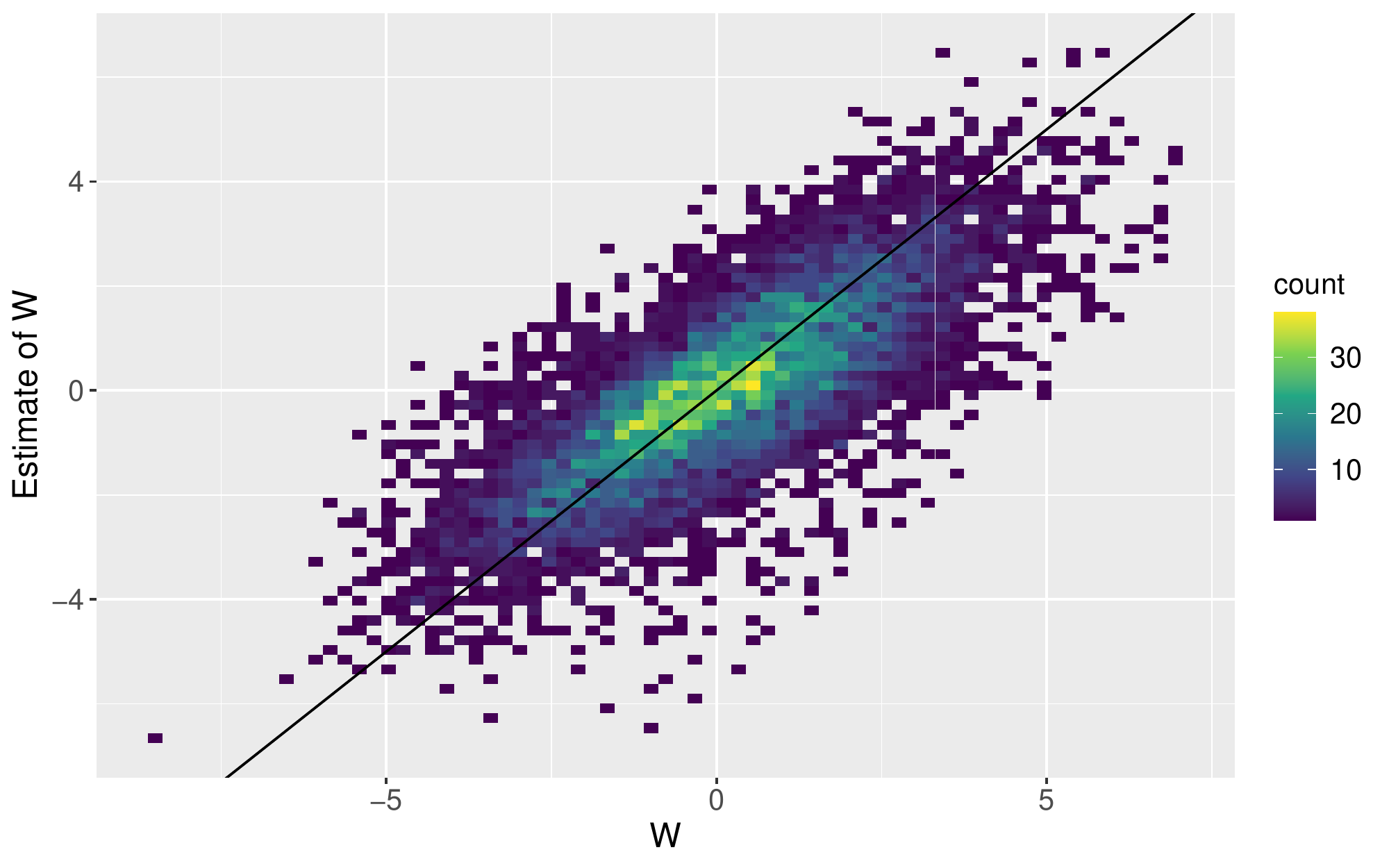}
		\caption{MDAGAR: low}\label{fig: wdl}
	\end{subfigure} 
	\begin{subfigure}[t]{0.3\textwidth}
		\centering
		\includegraphics[scale=0.25]{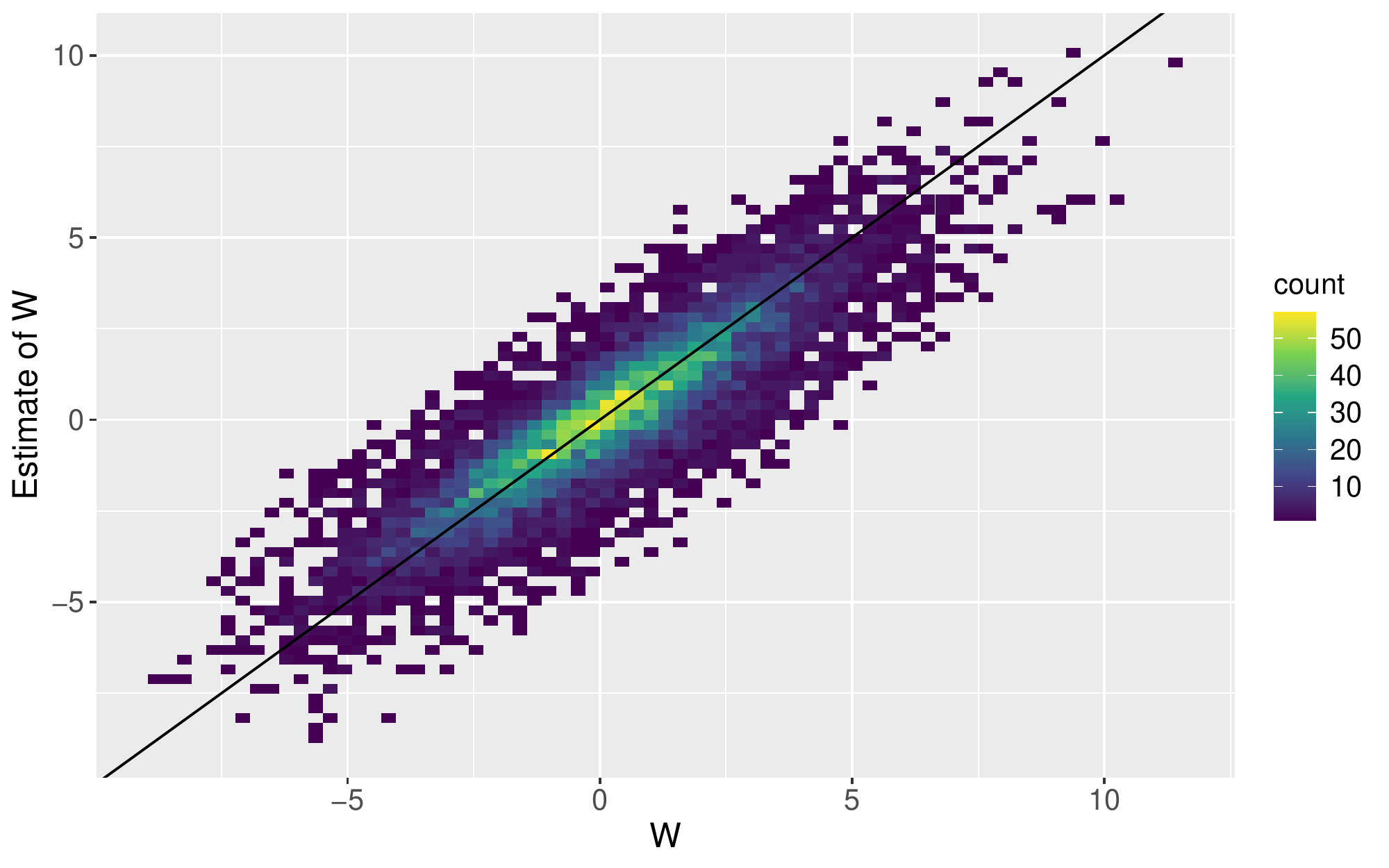}
		\caption{MDAGAR: medium}\label{fig: wdm}
	\end{subfigure} 
	\begin{subfigure}[t]{0.3\textwidth}
		\centering
		\includegraphics[scale=0.25]{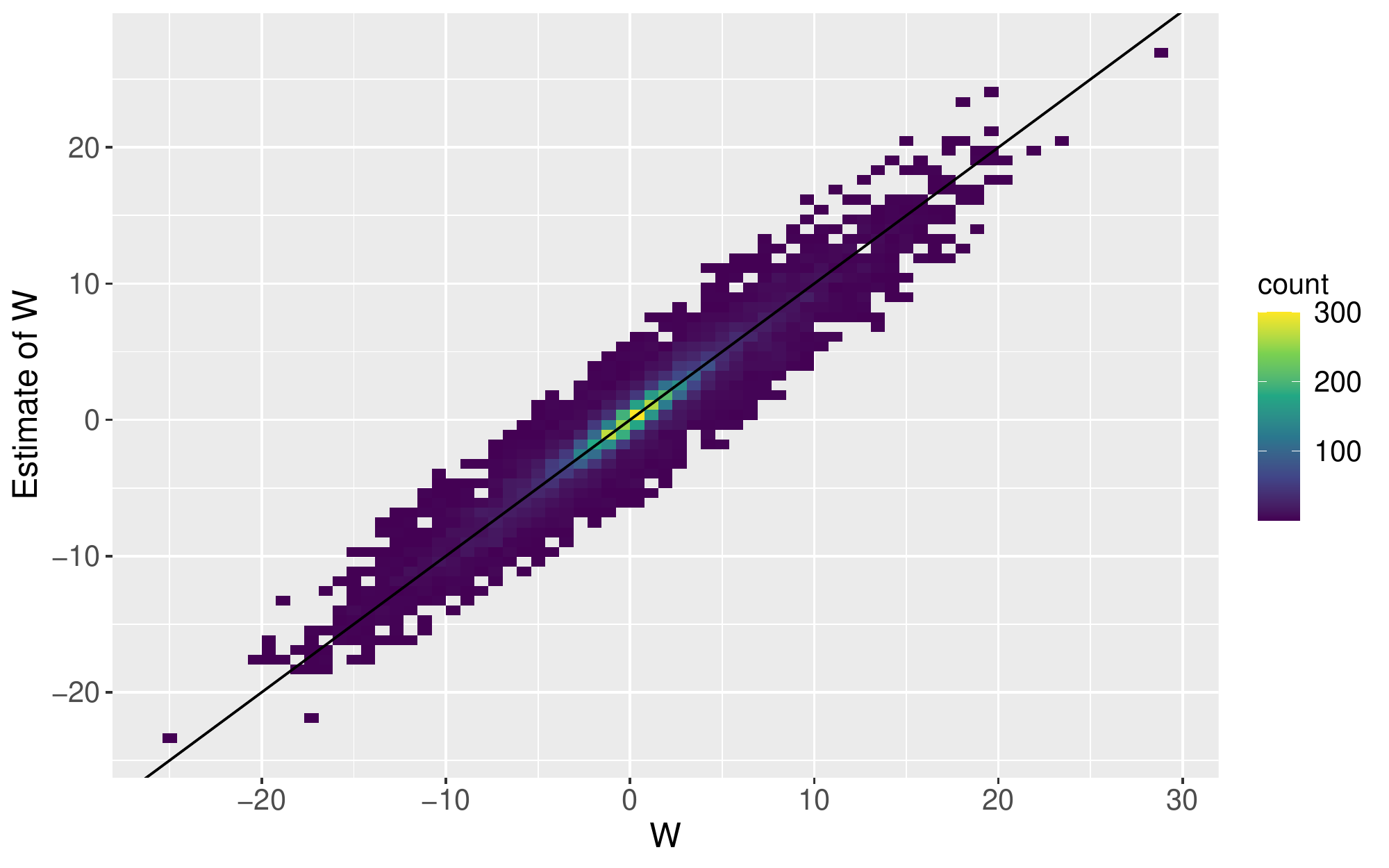}
		\caption{MDAGAR: high}\label{fig: wdh}
	\end{subfigure}
	\vfill
	\begin{subfigure}[t]{0.3\textwidth}
		\centering
		\includegraphics[scale=0.25]{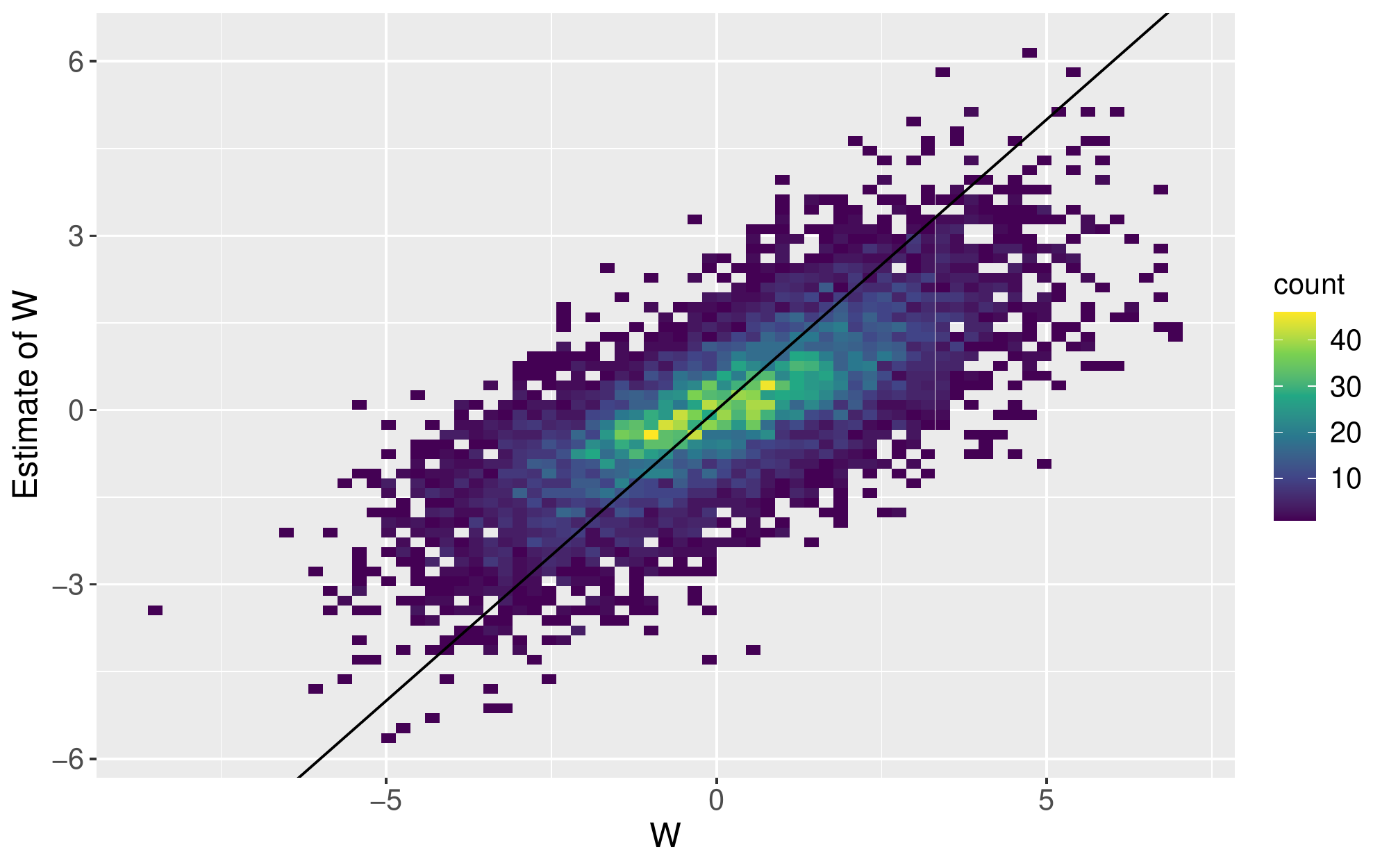}
		\caption{GMCAR: low}\label{fig: wcl}
	\end{subfigure} 
	\begin{subfigure}[t]{0.3\textwidth}
		\centering
		\includegraphics[scale=0.25]{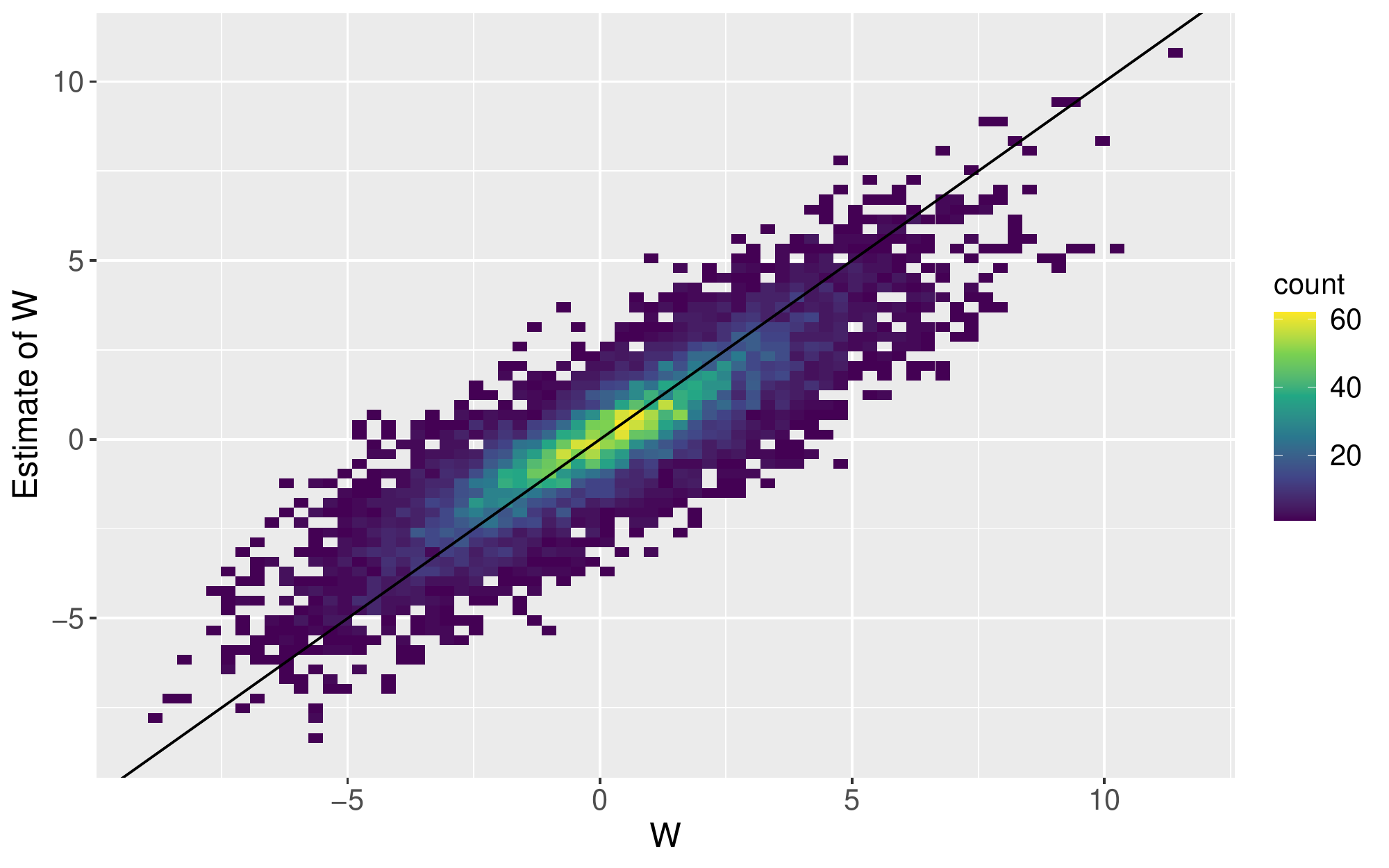}
		\caption{GMCAR: medium}\label{fig: wcm}
	\end{subfigure} 
	\begin{subfigure}[t]{0.3\textwidth}
		\centering
		\includegraphics[scale=0.25]{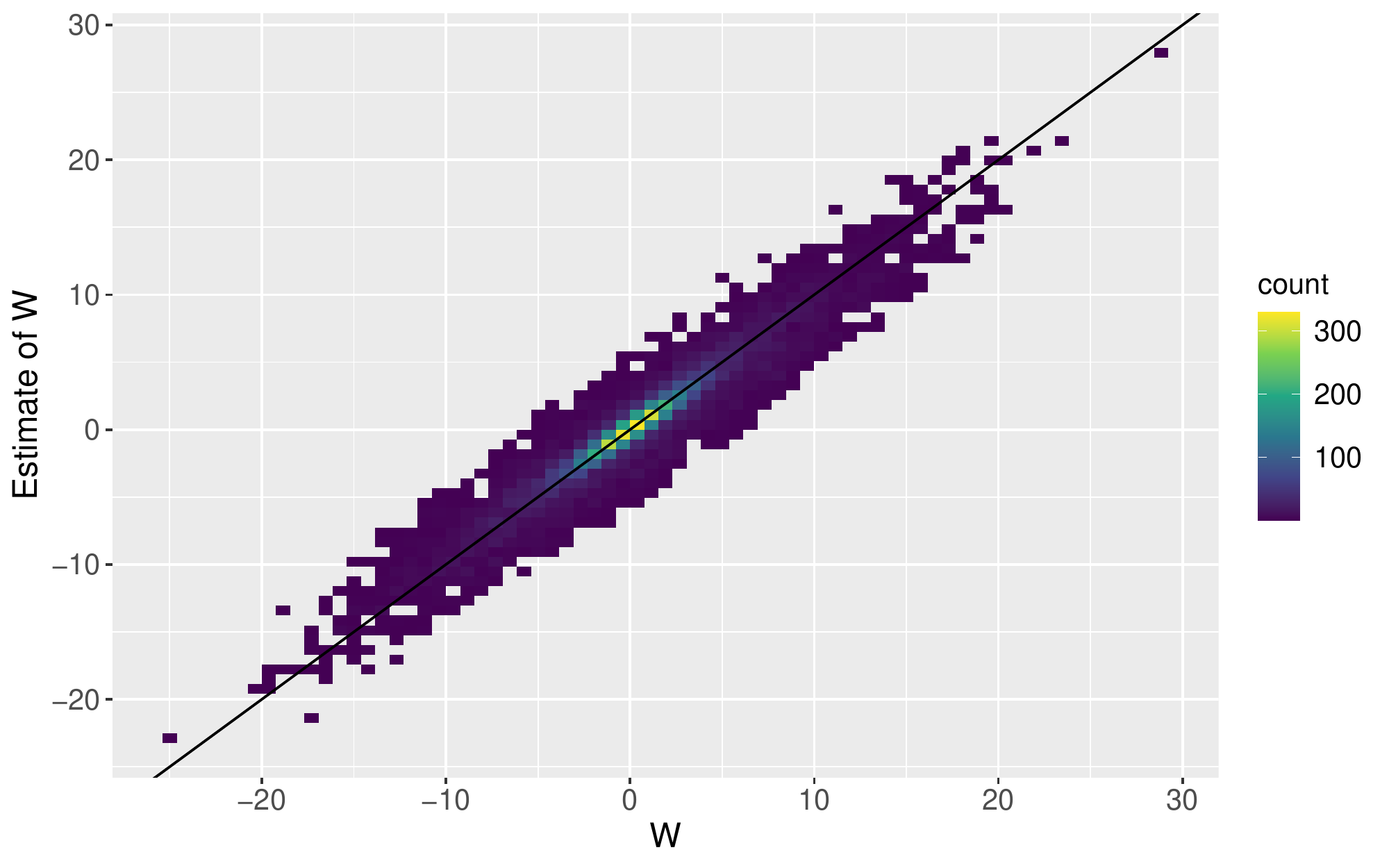}
		\caption{GMCAR: high}\label{fig: wch}
	\end{subfigure}
	\caption{Scatter plots for estimates of spatial random effects (y axis) against the true values (x axis) with $45^{\circ}$ lines over 85 datasets: (a)--(c) are estimates from MDAGAR model with low, medium and high correlation, while (d)--(f) are the corresponding estimates from GMCAR.}\label{fig: west}
	%		\end{adjustwidth}
\end{figure}

Figure~\ref{fig: west} presents scatter plots for the true values (x axis) of spatial random effects against their posterior estimates (y axis). To be precise, each panel plots $85 \times 48 \times 2 = 8160$ true values of the elements of the $96 \times 1$ vector $\bm{w}$ for $85$ datasets against their corresponding posterior estimates. We see strong agreements between the true values and their estimates for both MDAGAR and GMCAR. The agreement is more pronounced for the datasets corresponding to medium and high correlations. For the low-correlation datasets, the agreement is clearly weaker although MDAGAR does slightly better than GMCAR.

We compute 
$\displaystyle D_{KL}(N(\bm{0}, \bm{Q}_{true})|| N(\bm{0}, \bm{Q}_w)) = \frac{1}{2}\left[\log\left(\frac{\det(\bm{Q}_{true})}{\det(\bm{Q}_{w})}\right) + \mbox{tr}(\bm{Q}_{w}\bm{Q}_{true}^{-1}) - qk\right]$, which is the Kullback-Leibler Divergence between the model for $\bm{w}$ with the true generative precision matrix ($\bm{Q}_{true}$) and those with MDAGAR and GMCAR precisions ($\bm{Q}_w$). Using the posterior samples in the precision matrix, we evaluate the posterior probability that $D_{KL}(N(\bm{0}, \bm{Q}_{true})|| N(\bm{0}, \bm{Q}_{MDAGAR}))$ is smaller than $D_{KL}(N(\bm{0}, \bm{Q}_{true})|| N(\bm{0}, \bm{Q}_{w}))$. Figure~\ref{fig: KL} depicts a density plot of these probabilities over the 85 data sets. When correlations are low and medium, the MDAGAR has a mean probability of around $69\%$ to be closer to the true model than the GMCAR, while for high correlations GMCAR excels with an average probability of $72\%$ to be closer to the true model. These findings are consistent with the results of AMSE, where the GMCAR tended to perform better when the correlations are high. Additional comparative diagnostics from MDAGAR and GMCAR such as coverage probabilities for parameters and correlation between random effects for two diseases in the same state are presented in \ref{sup_cp}.

\subsection{Model selection for different disease orders}	
We now evaluate the effectiveness of the method in Section~\ref{selection} at selecting the model with the correct ordering of diseases. We used the \texttt{bridgesampling} package in \texttt{R} to compute $p(M_i\given \bm{y}^{(n)}) = \mbox{max}_{t=1,\ldots,6}\ p(M_t\given \bm{y}^{(n)})$ for each of $n=50\times 6$ data sets generated as described in Section~\ref{gen}. Table~\ref{tab:sim_model} presents the probability of each model being selected for different true model scenarios. The probability of selecting the true model is shown in bold along the diagonal. Our experiment reveals that bridge sampling is extremely effective at choosing the correct order. It was able to identify the correct order between $80\%$ to $ 90\%$, which is substantially larger than any of the probability of choosing any of the misspecified models. %Furthermore, for six models each pair out of the three pairs of models can be recognized as each other with probabilities at $10\% - 20\%$. In detail, $M_1$ and $M_3$ are a pair of models with the probability to be misclassified, i.e. the probability of $M_1$ recognized as $M_3$ is $0.1$ while the probability is $0.14$ in turn. The other two pairs are $M_2$ and $M_5$, $M_4$ and $M_6$. 

\begin{table}[htb]
	\centering
	\caption{Proportion of times ($\pi(M_i)$) bridge sampling chose the model with the correct order out of the 50 data sets with that order.}\label{tab:sim_model}
	\begin{tabular}{ccccccc}
		\hline
		True model & $\pi(M_1)$ & $\pi(M_2)$ & $\pi(M_3)$ & $\pi(M_4)$ & $\pi(M_5)$& $\pi(M_6)$ \\
		\hline
		$M_1$ & \textbf{0.90} & 0.00  & 0.10 & 0.00 & 0.00 & 0.00 \\
		$M_2$ & 0.00  & \textbf{0.86} & 0.00  & 0.00 & 0.14  & 0.00 \\
		$M_3$ & 0.14  & 0.00  & \textbf{0.86} & 0.00  & 0.00  & 0.00 \\
		$M_4$ & 0.00  & 0.00  & 0.00  & \textbf{0.90} & 0.00  & 0.10 \\
		$M_5$ & 0.00 & 0.22 & 0.00  & 0.00 & \textbf{0.78} & 0.00 \\
		$M_6$ & 0.00  & 0.00  & 0.00  & 0.16 & 0.00   & \textbf{0.84} \\ 
		\hline
	\end{tabular}
\end{table}

\section{Multiple Cancer Analysis from SEER}\label{data}
We now turn to an areal dataset with 4 different cancers using the MDAGAR model. The data set is extracted from the SEER$^*$Stat database using the SEER$^*$Stat statistical software \citep{seer}. The dataset consists of four cancers: lung, esophagus, larynx and colorectal, where the outcome is the 5-year average age-adjusted incidence rates (age-adjusted to the 2000 U.S. Standard Population) per 100,000 population in the years from 2012 to 2016 across 58 counties in California, USA, as mapped in Figure~\ref{fig: cancer_incidence_maps}. The maps exhibit preliminary evidence of correlation across space and among cancers. Cutoffs for the different levels of incidence rates are quantiles for each cancer. For all four cancers, incidence rates are relatively higher in counties concentrated in the middle northern areas including Shasta, Tehama, Glenn, Butte and Yuba than those other areas. In general, northern areas have higher incidence rates than in the southern part. This is especially pronounced for lung cancer and esphogus cancer. For larynx cancer, in spite of the highest incidence rate concentrated in the north, the incidence rates in the south are mostly at the same high level. For colorectal cancer, the edge areas at the bottom also exhibit high incidence rates. 

Overall, counties with similar levels of incidence rates tend to depict some spatial clustering. We analyze this data set using \eqref{eq: bivariate_DAGAR_bhm} with the following prior specification
\begin{align}\label{eq: priors}
p(\bm{\eta},\bm{\rho},\bm{\tau},\bm{\sigma},\bm{w}) &= \prod_{i=1}^q Unif(\rho_i\given 0,1)\times \prod_{i=2}^q\prod_{j=1}^{i-1}N(\bm{\eta}_{ij}\given 0, 0.01\bm{I}_2)\times \prod_{i=1}^q N(\bm{\beta}_i\given 0, 0.001\bm{I})\nonumber\\ 
&\times \prod_{i=1}^qIG(1/\tau_i\given 2,0.1)\times \prod_{i=1}^q IG(\sigma_i^2\given 2,1) \times N(\bm{w}\given \bm{0}, \bm{Q_w})\;.   
\end{align} 
%where $\bm{Q_w}$ is the precision matrix of $\bm{w}$. County attributes as well as a potential risk factor, adult cigarette smoking rates, are included in the regression as covariates in Section~\ref{data}. 
We also discuss a case excluding the risk factor (see Section~\ref{sup_case2}).

%\subsection{Case 1: Including cigarette smoking as a risk factor}\label{rf}
For covariates, we include county attributes that possibly affect the incidence rates,  including percentages of residents younger than 18 years old (young$_{ij}$), older than 65 years old (old$_{ij}$), with education level below high school (edu$_{ij}$), percentages of unemployed residents (unemp$_{ij}$), black residents (black$_{ij}$), male residents (male$_{ij}$), uninsured residents (uninsure$_{ij}$), and percentages of families below the poverty threshold (poverty$_{ij}$). All covariates are common for different cancers and extracted from the SEER$^*$Stat database \citep{seer} for the same period, 2012 - 2016. 
Since cigarette smoking is a common risk factor for cancers, adult smoking rates (smoke$_{ij}$) for 2014--2016 were obtained from the California Tobacco Facts and Figures 2018 database \citep{california2018california}. Spatial patterns in the map of adult cigarette smoking rates, shown in Figure~\ref{fig: covariates}, are similar to the incidence of cancers, especially lung and esophageal cancers, the highest smoking rates are concentrated in the north. While some central California counties (e.g., Stanislaus, Tuolumne, Merced, Mariposa, Fresno and Tulare) also exhibit high rates, although there is clearly less spatial clustering of the high rates than in the north. 

Since the order of cancers in the DAG specify the model, we fit all $4!=24$ models using \eqref{eq: bivariate_DAGAR_bhm} and compute the marginal likelihoods using bridge sampling (Section ~\ref{selection}). By setting the prior model probabilities as $p(M = M_t) = \frac{1}{24}$ for $t=1,2,\ldots,24$, we compute the posterior model probabilities using (\ref{eq:bma}). These are presented in Table~\ref{tab: post_prob}. We obtain Bayesian model averaged (BMA) estimates using (\ref{eq: ema}) with the weights in Table~\ref{tab: post_prob}. Among all models, model $M_{10}$ is selected as the best model with the largest posterior probability $0.577$ and the corresponding conditional structure is $[\mbox{esophageal}] \times [\mbox{larynx}\given\mbox{esophageal}] \times [\mbox{colorectal}\given \mbox{esophageal}, \mbox{larynx}] \times [\mbox{lung}\given \mbox{esophageal}, \mbox{larynx}, \mbox{colorectal}]$. %The interpretation of important parameters are based on the estimates from the best model.

%Furthermore, we will discuss the spatial associations for each cancer as well as associations among the cancers within each county. 
Table~\ref{tab:cov_smoke} is a summary of the parameter estimates including regression coefficients, spatial autocorrelation ($\rho_i$), spatial precision ($\tau_i$) and noise variance ($\sigma_i^2$) for each cancer. From $M_{10}$ and BMA, we find the regression slopes for the percentage of smokers and uninsured residents are significantly positive and negative, respectively, for esophageal cancer. The negative association between percentage of uninsured and esophageal cancer may seem surprising, but is likely a consequence of spatial confounding with counties exhibiting low incidence rates for esophageal cancer having a relatively large number of uninsured residents (see top right in \ref{fig: cancer_incidence_maps} and the right most figure in \ref{fig: covariates}). Since esophageal cancer has low incidence rates, this association could well be spurious due to spatial confounding. Percentage of smokers is, unsurprisingly, found to be a significant risk factor for lung cancer, while the percentage of blacks seems to be significantly associated with elevated incidence of larynx cancer. In addition, we tend to see that percentage of population below the poverty level has a pronounced association with higher rates of lung and esophageal cancer. 

Recall from Section~\ref{sec: CMDAGAR} that $\rho_1$ is the residual spatial autocorrelation for esophageal cancer after accounting for the explanatory variables, while $\rho_i$ for $i = 2, 3, 4$ are residual spatial autocorrelations after accounting for the explanatory variables and the preceding cancers in the model $M_{10}$. From Table~\ref{tab:cov_smoke} we see that esophageal cancer exihibits relatively weaker spatial autocorrelation, while the residual spatial autocorrelations for larynx and colorectal cancers after accounting for preceding cancers are both at moderate levels of around 0.5. Similarly for the spatial precision $\tau_i$,  larynx appears to have the smallest conditional variability while that for colorectal and lung are slightly larger. 

\begin{table}[H]
	\centering
	\caption{Posterior means (95$\%$ credible intervals) for parameters estimated from $M_{10}$ and BMA estimates for regression coefficients only for the SEER four cancer dataset.}\label{tab:cov_smoke}
	\hspace*{-1.5cm}{\small\begin{tabular}{cccccc}
			\hline
			Parameters& Model& Esophageal & Larynx & Colorectal & Lung \\
			\hline
			Intercept& $M_{10}$ & 16.76 (4.06, 29.56) & 6.37 (-1.16, 13.89) & 19.16 (-11.94, 49.72) & 28.68 (-18.3, 74.93) \\
			& BMA&15.87 (2.92, 28.63) & 6.85 (-0.71, 14.38) & 18.21 (-14.03, 49.07) & 28.25 (-18.12, 74.52)\\
			Smokers ($\%$) & $M_{10}$ &\textbf{0.25 (0.12, 0.37)} & 0.04 (-0.03, 0.12) & 0.23 (-0.12, 0.57) & \textbf{0.81 (0.08, 1.62)} \\
			& BMA&\textbf{0.23 (0.10, 0.36)} & 0.05 (-0.03, 0.12) & 0.22 (-0.13, 0.58) & \textbf{0.80 (0.08, 1.59)}\\
			Young ($\%$) & $M_{10}$ &-0.12 (-0.31, 0.07) & -0.07 (-0.18, 0.04) & 0.27 (-0.2, 0.76) & -0.08 (-0.90, 0.74) \\
			& BMA&-0.11 (-0.3, 0.08) & -0.08 (-0.19, 0.03) & 0.29 (-0.18, 0.78) & -0.01 (-0.86, 0.82)\\
			Old ($\%$)  & $M_{10}$ &-0.11 (-0.25, 0.04) & -0.05 (-0.14, 0.03) & 0.10 (-0.28, 0.48) & -0.09 (-0.81, 0.67) \\
			& BMA&-0.10 (-0.25, 0.05) & -0.05 (-0.14, 0.03) & 0.10 (-0.29, 0.49) & -0.08 (-0.79, 0.66)\\
			Edu ($\%$)  & $M_{10}$ &0.02 (-0.08, 0.12) & -0.02 (-0.08, 0.04) & 0.16 (-0.12, 0.43) & -0.20 (-0.75, 0.31) \\
			& BMA&0.02 (-0.09, 0.12) & -0.02 (-0.07, 0.04) & 0.15 (-0.14, 0.42) & -0.24 (-0.79, 0.27)\\
			Unemp ($\%$) & $M_{10}$ &-0.13 (-0.29, 0.03) & 0.01 (-0.08, 0.10) & -0.09 (-0.54, 0.37) & 0.60 (-0.47, 1.55) \\
			& BMA&-0.12 (-0.28, 0.05) & 0.01 (-0.08, 0.1) & -0.08 (-0.54, 0.38) & 0.61 (-0.43, 1.56) \\
			Black ($\%$) & $M_{10}$ &0.14 (-0.06, 0.34) & \textbf{0.14 (0.03, 0.26)} & -0.16 (-0.73, 0.39) & 0.15 (-1.06, 1.29) \\
			& BMA&0.13 (-0.07, 0.33) & \textbf{0.15 (0.03, 0.27)} & -0.18 (-0.75, 0.39) & 0.14 (-1.02, 1.25)\\
			Male ($\%$)  & $M_{10}$ &-0.04 (-0.17, 0.09) & 0.00 (-0.07, 0.08) & 0.24 (-0.12, 0.60) & 0.14 (-0.57, 0.79) \\
			& BMA&-0.04 (-0.17, 0.09) & 0 (-0.07, 0.08) & 0.24 (-0.12, 0.62) & 0.14 (-0.55, 0.82)\\
			Uninsured ($\%$) & $M_{10}$ &\textbf{-0.24 (-0.44, -0.04)} & -0.08 (-0.20, 0.04) & 0.07 (-0.44, 0.58) & 0.01 (-0.82, 0.86) \\
			& BMA&\textbf{-0.23 (-0.42, -0.02)} & -0.08 (-0.2, 0.04) & 0.09 (-0.42, 0.61) & 0 (-0.81, 0.82)\\
			Poverty ($\%$) & $M_{10}$ &0.30 (-0.24, 0.84) & 0.20 (-0.12, 0.51) & -0.06 (-1.51, 1.45) & 0.85 (-2.15, 3.85) \\
			& BMA&0.32 (-0.23, 0.87) & 0.2 (-0.12, 0.51) & -0.08 (-1.54, 1.42) & 0.8 (-2.14, 3.75)\\
			\hline
			$\rho_i$ &$M_{10}$ & 0.25 (0.01, 1.00) & 0.33 (0.01, 0.96) & 0.50 (0.03, 0.97) & 0.52 (0.03, 0.99) \\
			$\tau_i$ & $M_{10}$ &25.27 (5.08, 61.57) & 27.60 (8.05, 60.42) & 19.97 (3.06, 55.61) & 20.31 (1.77, 55.92) \\
			$\sigma_{i}^2$ &$M_{10}$ & 1.67 (1.11, 2.47) & 0.49 (0.28, 0.75) & 8.22 (1.09, 14.23) & 1.19 (0.18, 5.21)\\
			\hline
	\end{tabular}}
\end{table}

\begin{figure}
	\centering
	\begin{subfigure}{0.5\linewidth}
		\centering
		\includegraphics[width=72mm]{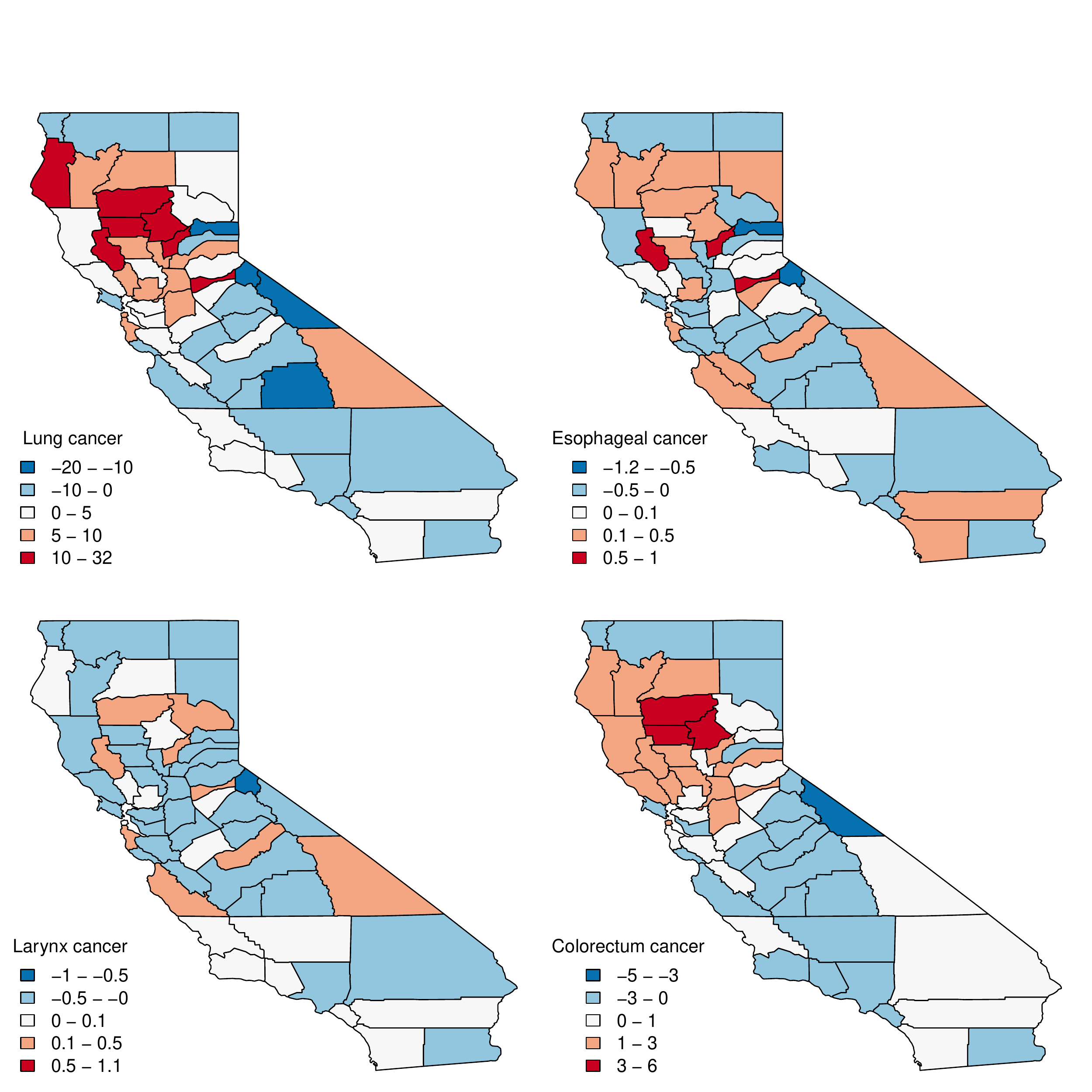}
		\caption{BMA: posterior mean spatial random effects}\label{fig: random_covariates_ave}
	\end{subfigure}
	\hfill
	\hskip -6.5cm\begin{subfigure}{0.5\linewidth}
		\centering
		\includegraphics[width=72mm]{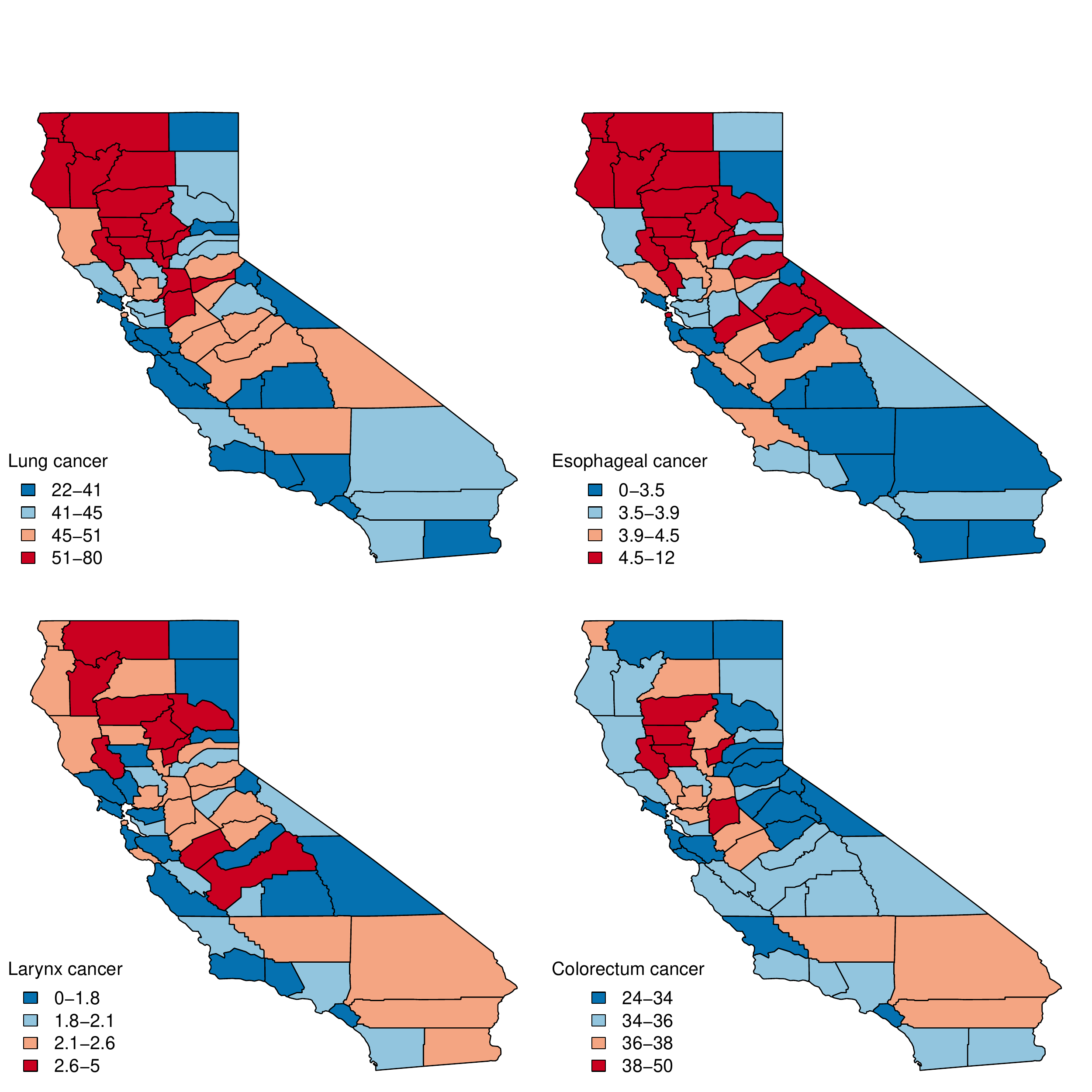}
		\caption{BMA: posterior mean incidence rates}\label{fig: post_incidence_maps_ave}
	\end{subfigure}
	\vfill
	\begin{subfigure}{0.5\linewidth}
		\centering
		\includegraphics[width=72mm]{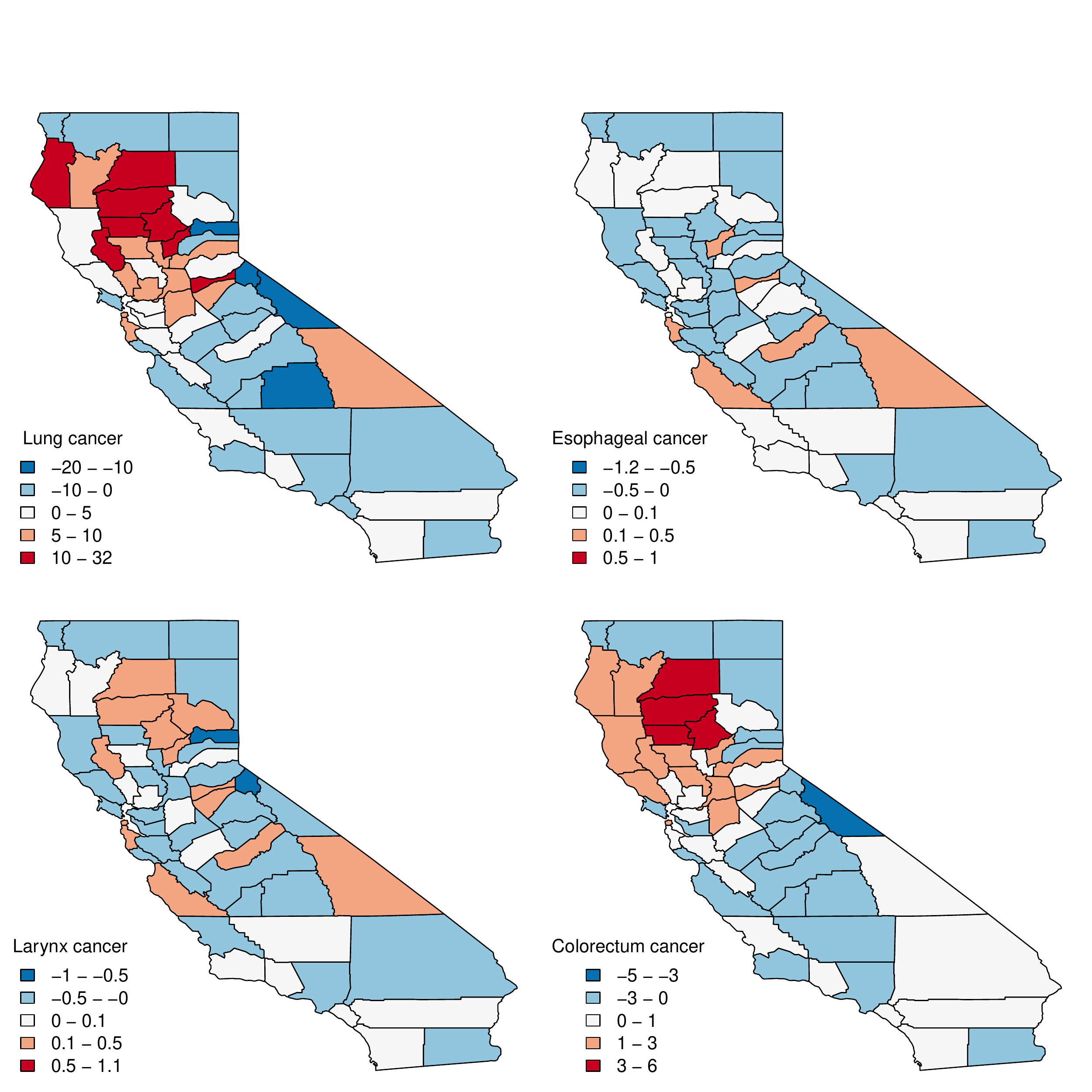}
		\caption{$M_{10}$: posterior mean spatial random effects}\label{fig: random_covariates}
	\end{subfigure}
	\hfill
	\hskip -6.5cm\begin{subfigure}{0.5\linewidth}
		\centering
		\includegraphics[width=72mm]{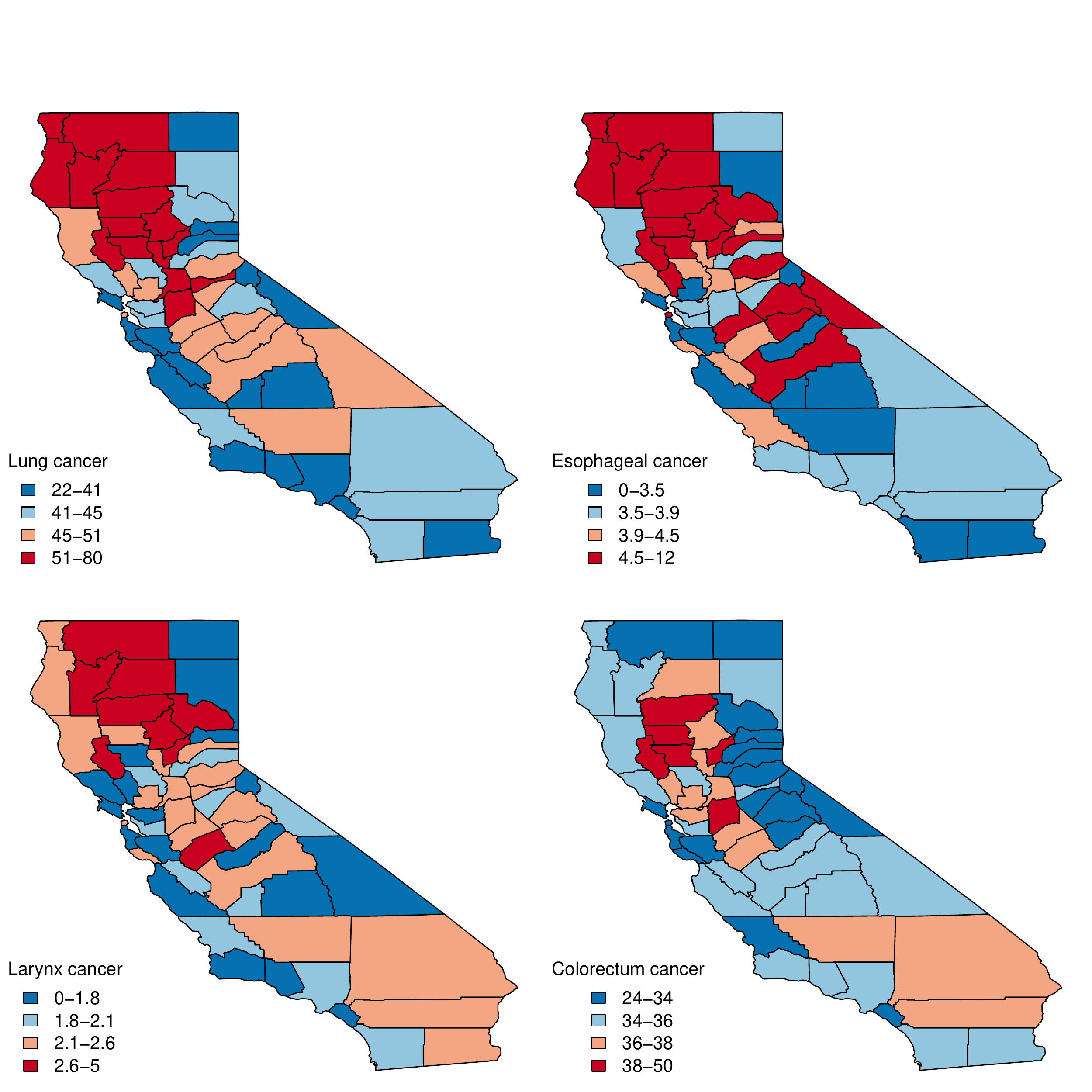}
		\caption{$M_{10}$: posterior mean incidence rates}\label{fig: post_incidence_maps}
	\end{subfigure}
	\caption{Maps of posterior results using BMA and the highest probabilty model $M_{10}$ for lung, esophagus, larynx and colorectal cancer in California including posterior mean spatial random effects and posterior mean incidence rates as shown in (a) (b) for BMA and (c) (d) for $M_{10}$; }\label{fig: postest}
\end{figure}

For the posterior mean incidence rates and spatial random effects $w_{ij}$, we present estimates from model $M_{10}$ and BMA. Figure \ref{fig: random_covariates_ave} and \ref{fig: post_incidence_maps_ave} are maps of posterior mean spatial random effects and model fitted incidence rates for four cancers obtained from BMA, while Figure \ref{fig: random_covariates} and \ref{fig: post_incidence_maps} show maps of those from model $M_{10}$. The posterior mean incidence rates from BMA and $M_{10}$ are in accord with each other, and both present DAGAR-smoothed versions of the original patterns in Figure~\ref{fig: cancer_incidence_maps}. For posterior means of spatial random effects, in general, the estimates from $M_{10}$ are similar to model averaged estimates, especially for lung and colorectal cancers, exhibiting relatively large positive values in the northern counties, where the incidence rates are high. However, for esophageal and larynx cancers we see slight discrepancies between $M_{10}$ and BMA in the north. The BMA estimates produce larger positive random effects, ranging between $0.1-0.5$, in most counties, while $M_{10}$ produces estimates between $0-0.1$ for esophageal cancer. More counties with random effects larger than $0.1$ are estimated from $M_{10}$ for larynx cancer. We believe this is attributable, at least in part, to another competitive model, $M_{15}=[\mbox{larynx}] \times [\mbox{esophagus}\given\mbox{larynx}] \times [\mbox{lung}\given \mbox{larynx}, \mbox{esophagus}] \times [\mbox{colorectal}\given \mbox{larynx}, \mbox{esophagus}, \mbox{lung}]$ (posterior probability $0.342$), which contributes to the BMA. On the other hand, the effects of some important county-level covariates play an essential role in the discrepancy between the estimates of random effects and model fitted incidence rates for each cancer.

Recall from Section~\ref{Methods} that $\eta_{0ii'}$ and $\eta_{1ii'}$ reflect the associations among cancers that can be attributed to spatial structure. Specifically, larger values of $\eta_{0ii'}$ will indicate inherent associations unrelated to spatial structure, while the magnitude of $\eta_{1ii'}$ reflects associations due to spatial structure. Figure~\ref{fig: eta_final_new} presents posterior distributions of $\bm{\eta}$ for all pairs of cancers. We see from the distribution of $\eta_{043}$ that there is a pronounced non-spatial component in the association between lung and colorectal cancers. Similar, albeit somewhat less pronounced, non-spatial associations are seen between larynx and esophageal cancers and between lung and larynx cancers. Analogously, the posterior distributions for $\eta_{143}$ and $\eta_{132}$ tend to have substantial positive support suggesting substantial spatial cross-correlations between lung and colorectal cancers and between colorectal and larynx cancers. Interestingly, we find negative support in the posterior distributions for $\eta_{121}$ and $\eta_{142}$. The negative mass implies that the covariance among cancers within a region is suppressed by strong dependence with neighboring regions. This seems to be the case for associations between lung and esophageal cancers and between lung and larynx cancers. 

We also present supplementary analysis that excludes adult smoking rates from the covariates, which we refer to as ``Case 2''. Figure~\ref{fig: correlation_cases} shows estimated correlations between pairwise cancers in each of the 58 counties. The top row presents the correlations including smoking rates (``Case 1'') as has been analyzed here. The bottom row presents the corresponding maps for ``Case 2''. Interestingly, accounting for smoking rates substantially diminishes the associations among esophageal, colorectal and lung cancers. These are significantly associated in ``Case 2'' but only lung and colorectal retain their significance after accounting for smoking rates.

%\begin{figure}[H]
%	\centering
%	\includegraphics[width=\linewidth]{estimates_cancerfinal.pdf}
%	\caption{Maps of posterior mean incidence rates for lung, esophagus, larynx, and colorectum cancer in California}
%\end{figure}

% Table generated by Excel2LaTeX from sheet 'table_est_new'

\section{Discussion}
\label{discussion}
We have developed a conditional multivariate ``MDAGAR'' model to estimate spatial correlations for multiple correlated diseases based on a currently proposed class of DAGAR models for univariate disease mapping, as well as providing better interpretation of the association among diseases. We demonstrate that MDAGAR tends to outperform GMCAR when association between spatial random effects for different diseases is weak or moderate. Inference is competitive when associations are strong. MDAGAR retains the interpretability of spatial autocorrelations, as in univariate DAGAR, separating the spatial correlation for each disease from any inherent or endemic association among diseases. While MDAGAR, like all DAG based models, is specified according to a fixed order of the diseases, we show that a bridge sampling algorithm can effectively choose among the different orders and also provide Bayesian model averaged inference in a computationally efficient manner. 

Our data analysis reveals that correlations between incidence rates for different cancers are impacted by covariates. For example, eliminating adult cigarette smoking rates produces similar spatial patterns for the incidence rates of esophageal, lung and colorectal cancer. In addition, the significant correlation between lung and esophageal cancer, even after accounting for smoking rates, implies other inherent or endemic association such as latent risk factors and metabolic mechanisms. We also see that the MDAGAR based posterior estimates of the latent spatial effects in Figure~\ref{fig: random_covariates_ave} and \ref{fig: random_covariates} resemble those from an MDAGAR without covariates (Figure~\ref{fig: random_nocovariates}), while the maps for the estimated incidence rates in Figure~\ref{fig: post_incidence_maps_ave} and \ref{fig: post_incidence_maps} account for the spatial variability of the covariates.

%For more discussion about spatial random effects and fixed effects from covariates, we also fit the model in the same order ($M_{10}$) as Case 1 but with no covariates (only with intercept) in regression and generate maps of posterior mean spatial random effects (see Web Appendix B Figure~\ref{fig: random_nocovariates}) for comparison. When no covariates is involved in regression, the distribution of spatial random effects are consistent with the incidence of cancers in general. As expected, overall, spatial random effects become smaller after accounting for fixed effects of all explanatory covariates given signicificant positive effects. In detail, for lung and esophageal cancers, given that smoking rates take major effects, positive spatial random effects are still concentrated in northen counties with much higher values for lung cancer, while negative effects for counties in the middle central counties offset the large fixed effects from high smoking rates. For larynx cancer, clusters with higher incidence rates in middle and south area are likely to partially associate with the significant positive effect of black given larger percentage of black residents in those clusters as shown in Figure~\ref{fig: covariates}, and the high incidence rates in the top area present in the larger spatial random effects in same counties. Since no covariate has significant effect on the incidence of colorectum cancer, spatial random effects in Figure~\ref{fig: random_covariates} only decreases a little bit with similar distribution as the one without covariates.

Future challenges will include scalability with very large number of diseases because, as we have seen, the number of models to be fitted grows exponentially with the number of diseases. One way to obviate this issue is to adopt a joint modeling approach analogous to order-free MCAR models \citep{jin2007order} that build rich spatial structures from linear transformations of simpler latent variables. For instance, we can develop alternate MDAGAR models by specifying $\bm{w} = \bm{\Lambda} \bm{f}$, where $\bm{\Lambda}$ is a suitably specified matrix and $\bm{f}$ is a latent vector whose components follow independent univariate DAGAR distributions. This will avoid the order dependence, but the issue of identifying and specifying $\bm{\Lambda}$ will need to be considered as will the interpretation of disease specific spatial autocorrelations.

\section*{Acknowledgements}
The work of the first and third authors have been supported in part by the Division of Mathematical Sciences (DMS) of the National Science Foundation (NSF) under grant 1916349 and by the National Institute of Environmental Health Sciences (NIEHS) under grants R01ES030210 and 5R01ES027027. The work of the second author was supported by the Division of Mathematical Sciences (DMS) of the National Science Foundation (NSF) under grant 1915803.

\section*{Supplementary Materials}\label{sec: SM}
The R code implementing our models are available at \url{https://github.com/LeiwenG/Multivariate_DAGAR}.

\bibliographystyle{biom}
\bibliography{Banerjee.bib}

\renewcommand{\theequation}{S.\arabic{equation}}
\renewcommand{\thesection}{S.\arabic{section}}
\renewcommand{\thetable}{S.\arabic{table}}
\renewcommand{\thefigure}{S.\arabic{figure}}

\section{Algorithm for Model Implementation}
\label{sup_implementation}
We outline model implementation for \eqref{eq: bivariate_DAGAR_bhm} using Markov Chain Monte Carlo (MCMC). We update $\{\bm{w}, \bm{\beta}, \bm{\sigma}, \bm{\tau}, \bm{\eta}_2,\dots, \bm{\eta}_q\}$ using Gibbs steps, while the elements of $\bm{\rho}$ are updated from their full conditional distributions using Metropolis random walk steps \citep{robert2004monte}. A particularly appealing feature of our proposed MDAGAR model is that the spatial weight parameters $\bm{\eta} = \{\bm{\eta}_2,\dots, \bm{\eta}_q\}$ render Gaussian full conditional distributions in addition to the customary Gaussian full conditional distributions for $\bm{\beta}$ and $\bm{w}$. As a matter of notational convenience for the derivations that follow, we use $N(\bm{\mu}, \bm{V})$ to denote the normal distribution with variance-covariance matrix $\bm{V}$. This difference from our notation in the main manuscript where we use the precision matrix in the argument of normal distribution.

\subsection{Full Conditional Distributions}
The full conditional distribution for each $\bm{\beta}_i$ is
\begin{align}
\bm{\beta}_i | \bm{y}_i, \bm{w}_i, \sigma_{i}^2 \sim N(\bm{M}_i\bm{m}_i, \bm{M}_i)\label{eq:beta}
\end{align}
where $\bm{M}_i = \left(\frac{1}{\sigma_{i}^2}\bm{X}_i^\top\bm{X}_i + \frac{1}{\sigma_{\beta}^2}\bm{I}_{p_i}\right)^{-1}$ and $\bm{m}_i = \frac{1}{\sigma_{i}^2}\bm{X}_i^\top\left(\bm{y}_i - \bm{w}_i\right)$. Similarly, the full conditional distribution of each $\sigma_{i}^2$ follows an inverse gamma distribution,
\begin{align}
\sigma_{i}^2 | \bm{y}_i, \bm{\beta}_i, \bm{w}_i \sim IG\left(a_\sigma+\frac{k}{2}, b_\sigma+\frac{1}{2}\left(\bm{y}_i - \bm{X}_i\bm{\beta}_i - \bm{w}_i\right)^\top\left(\bm{y}_i - \bm{X}_i\bm{\beta}_i - \bm{w}_i\right)\right).\label{eq:sigma}
\end{align}
The full conditional distribution for $\bm{w}_i$ for each $i = 2, \dots, q-1$ is
\begin{align}
&p\left(\bm{w}_i \given \bm{w}_1, \dots,\bm{w}_{i-1}, \bm{w}_{i+1}, \bm{w}_{i+1},\dots,\bm{w}_{q}, \bm{y}_i, \bm{\beta}_i, \sigma_{i}^2, \bm{\eta}_i, \dots,\bm{\eta}_q,\rho_i, \dots,\rho_q,\tau_i, \dots, \tau_q\right) \nonumber\\
&\quad\quad\quad\quad\quad\propto \prod_{n=i}^q\exp{\left\{-\frac{\tau_n}{2}\left(\bm{w}_n-\sum_{i'=1}^{n-1}\bm{A}_{ni'}\bm{w}_{i'}\right)^\top \bm{Q}\left(\rho_n\right)\left(\bm{w}_n-\sum_{i'=1}^{n-1}\bm{A}_{ni'}\bm{w}_{i'}\right)\right\}}\nonumber\\
&
\quad\quad\quad\quad\quad\quad\quad\quad\quad \times \exp{\left\{-\frac{1}{2\sigma_{i}^2}\left(\bm{y}_i-\bm{X}_i\bm{\beta}_i-\bm{w}_i\right)^\top\left(\bm{y}_i-\bm{X}_i\bm{\beta}_i-\bm{w}_i\right)\right\}}
\end{align}
which is equal to $N(\bm{w}_i \given \bm{G}_i\bm{g}_i, \bm{G}_i)$, where 
\begin{align*}
\bm{G}_i = \bigg[\tau_i\bm{Q}(\rho_i) + \sum_{n=i+1}^q\tau_n\bm{A}_{ni}^\top\bm{Q}(\rho_n)\bm{A}_{ni} + \frac{1}{\sigma_{i}^2}\bm{I}_k\bigg]^{-1}
\end{align*}
and $\displaystyle \bm{g}_i = \tau_i\bm{Q}(\rho_i)\sum_{n=1}^{i-1}\bm{A}_{in}\bm{w}_n + \sum_{n=i+1}^q\tau_n\bm{A}_{ni}^\top\bm{Q}(\rho_n)\left(\bm{w}_n-\sum_{i'=1, i' \neq i}^{n-1} \bm{A}_{ni'}\bm{w}_{i'}\right) + \frac{1}{\sigma_{i}^2}(\bm{y}_i - \bm{X}_i\bm{\beta}_i)$.
For $i=1$ and $q$, we have 
\begin{align*}
\bm{w}_1 | \bm{w}_2, \dots,\bm{w}_{q}, \bm{y}_1, \bm{\beta}_1, \sigma_{1}^2, \bm{\eta},\bm{\rho},\bm{\tau} &\sim N(\bm{G}_1\bm{g}_1, \bm{G}_1)\\
\bm{w}_q | \bm{w}_1, \dots, \bm{w}_{q-1}, \bm{y}_q, \bm{\beta}_q,\sigma_{1}^2, \bm{\eta}_q,\rho_q,\tau_q &\sim N(\bm{G}_q\bm{g}_q, \bm{G}_q)
\end{align*}, where
\begin{align*}
\bm{G}_1 &= \left(\tau_1\bm{Q}(\rho_1) +\sum_{n=2}^q\tau_n\bm{A}_{n1}^\top\bm{Q}(\rho_n)\bm{A}_{n1} + \frac{1}{\sigma_{1}^2}\bm{I}_k\right)^{-1},\\ 
\bm{g}_1 &= \tau_2\bm{A}_{21}^\top\bm{Q}(\rho_2)\bm{w}_2+\sum_{n=3}^q\tau_n\bm{A}_{n1}^\top\bm{Q}(\rho_n)\left(\bm{w}_n-\sum_{i'=2}^{n-1} \bm{A}_{ni'}\bm{w}_{i'}\right) + \frac{1}{\sigma_{1}^2}(\bm{y}_1 - \bm{X}_1\bm{\beta}_1),\\
\bm{G}_q &= \left(\tau_q\bm{Q}(\rho_q) + \frac{1}{\sigma_{q}^2}\bm{I}_k\right)^{-1}\\
\bm{g}_q &= \tau_q\bm{Q}(\rho_q)\sum_{n=1}^{q-1}\bm{A}_{qn}\bm{w}_n + \frac{1}{\sigma_{q}^2}(\bm{y}_q - \bm{X}_q\bm{\beta}_q).
\end{align*}
\noindent The full conditional distribution of each $\tau_i$ is
\begin{multline}
\tau_1 | \bm{w}_1, \rho_1 \sim G\left(a_{\tau_1}+\frac{k}{2}, b_{\tau_1}+\frac{1}{2}\bm{w}_1^T\bm{Q}(\rho_1)\bm{w}_1\right),\nonumber\\
\tau_i | \bm{w}_1, \dots, \bm{w}_i, \bm{\eta}_{i}, \rho_i\sim G\left(a_{\tau_i}+\frac{k}{2}, b_{\tau_i}+\frac{1}{2}\left(\bm{w}_i-\sum_{i'=1}^{i-1}\bm{A}_{i,i'}\bm{w}_{i'}\right)^\top\bm{Q}(\rho_i)\left(\bm{w}_i-\sum_{i'=1}^{i-1}\bm{A}_{i,i'}\bm{w}_{i'}\right)\right),\\ i = 2, 3, \dots, q
\end{multline}

We now derive the full conditional distributions for the $\bm{\eta}_i$s. From \eqref{eq:w_q} with $i=2$, each element in $\bm{w}_2$ can be written as $w_{2j} = \eta_{021}w_{1j} + \eta_{121}\sum_{j' \sim j}w_{1j'} + \epsilon_{2j}$, where $\epsilon_{2j}$ is the $j$th element in $\bm{\epsilon}_2$. To extract $\bm{\eta}_{21} = (\eta_{021}, \eta_{121})^\top$ from the matrix $\bm{A}_{21}$, $\bm{A}_{21}\bm{w}_1$ is rewritten as $\bm{Z}_{1}\bm{\eta}_{21}$ where $\bm{Z}_{1} = (\bm{w}_1, \bm{\zeta}_1)$ and $\bm{\zeta}_1 = \left(\sum_{j' \sim 1}w_{1j'}, \dots, \sum_{j' \sim k}w_{1j'}\right)^\top$. In general, $\bm{A}_{ii'}\bm{w}_{i'} = \bm{Z}_{i'}\bm{\eta}_{ii'}$ with $\bm{Z}_{i'} = (\bm{w}_{i'}, \bm{\zeta}_{i'})$, where $\bm{\zeta}_{i'} = \left(\sum_{j' \sim 1}w_{i'j'}, \dots, \sum_{j' \sim k}w_{i'j'}\right)^\top$. Consequently, (5) can be written as $\bm{w}_i = \bm{\delta}_i\bm{\eta}_i + \bm{\epsilon}_i$, where block matrix $\bm{\delta}_i = (\bm{Z}_{1}, \dots, \bm{Z}_{i-1})$. If $\bm{\eta}_{i} \sim N(\bm{\mu}_i, \bm{V}_i)$, then the full conditional distribution of $\bm{\eta}_{i}$ is
\begin{align}
p(\bm{\eta}_i \given \bm{w}_{1}, \dots, \bm{w}_{i}, \rho_i) &\propto \exp{\left\{-\frac{\tau_i}{2}(\bm{w}_i-\bm{\delta}_i\bm{\eta}_i)^\top \bm{Q}(\rho_i)(\bm{w}_i-\bm{\delta}_i\bm{\eta}_i)\right\}}\nonumber\\
&\quad \quad \quad \times \exp{\left\{-\frac{1}{2}(\bm{\eta}_i-\bm{\mu}_i)^\top\bm{V}_i^{-1}(\bm{\eta}_i-\bm{\mu}_i)\right\}}.\label{eq:psi}
\end{align}
The above is equal to $N(\bm{\eta}_i \given \bm{H}_i\bm{h}_i, \bm{H}_i)$, where $\bm{H}_i = \left(\tau_i\bm{\delta}_{i}^\top\bm{Q}(\rho_i)\bm{\delta}_{i} + \bm{V}_i^{-1}\right)^{-1}$ and $\bm{h}_i = \tau_i\bm{\delta}_{i}^\top\bm{Q}(\rho_i)\bm{w}_i + \bm{V}_i^{-1}\bm{\mu}_i$. For our analysis we set $\bm{\mu}_i = \bm{0}$ and $\bm{V}_i = 1000\bm{I}$.

\subsection{Metropolis within Gibbs}
Let $\gamma_i = \log(\frac{\rho_i}{1-\rho_i})$, $\gamma_i \in \mathbb{R}$ and $\bm{\gamma} = (\gamma_1, \dots, \gamma_q)^\top$. The full conditional distribution of $\bm{\gamma}$ is
\begin{align}
p(\bm{\gamma}|\bm{w}, \bm{\eta}_{2},\dots,\bm{\eta}_{q}, \bm{\tau}) \propto p(\bm{w} |\bm{\tau}, \bm{\eta}_2,\dots, \bm{\eta}_q, \bm{\rho}) \times p(\bm{\rho})|J|, \label{eq:gamma}
\end{align} 
where $p(\bm{w} |\bm{\tau}, \bm{\eta}_2,\dots, \bm{\eta}_q, \bm{\rho}) = N(\bm{w} \given \bm{G}\bm{g}, \bm{G})$, $\bm{G} = \left(\bm{Q}_w + \bm{\Sigma}^{-1}\right)^{-1}$, $\bm{g} =\bm{\Sigma}^{-1}\left(\bm{y} - \bm{X}\bm{\beta}\right)$, $\bm{\Sigma} = diag(\bm{\sigma})\otimes \bm{I}_k$ and $J = \prod_{i=1}^q\rho_i(1-\rho_i)$. Using the formula of transformation, $p(\bm{\rho})|J|$ is the prior for $\bm{\gamma}$ and in the right-hand side, $\bm{\rho}$ can be substituted by $\bm{\gamma}$ given $\rho_i = \frac{e^{\gamma_i}}{1+e^{\gamma_i}}$.

%Since $\bm{\rho}$ is embedded in the joint precision matrix $\bm{Q}_w$, the closed form of the full conditional distribution does not exist. Hence, at each Gibbs iteration, we generate one sample of $\bm{\rho}$ randomly using standard normal for each $\rho_i$ and decide whether to accept it depending on $N(\bm{w} | \bm{0}, \bm{Q}_w) \times p(\bm{\rho})$ given current values of $\bm{w}, \bm{\tau}, \bm{\eta}_2,\dots, \bm{\eta}_q$ in a Metropolis step. For each iteration $\ell = 1, \dots, L$ of the MCMC, we get the updates of $\bm{w}_i^{(\ell)}$, $\bm{\beta}_i^{(\ell)}$, $\sigma_{i}^{2(\ell)}$, $\tau_i^{(\ell)}$, and $\bm{\eta}_{i}^{(\ell)}$ using Gibbs sampler in terms of \eqref{eq:beta} - \eqref{eq:psi}, and update $\bm{\rho}^{(\ell)}$ in metropolis algorithm given these current values of parameters. Since the metropolis algorithm generates samples using random walk on the $\mathbb{R}$ domain, $\rho_1,\dots, \rho_q \in (0, 1)$ are transfered to logit scale, i.e. $\gamma_i = \log(\frac{\rho_i}{1-\rho_i})$, $\gamma_i \in \mathbb{R}$. The full posterior distribution of $\bm{\gamma} = (\gamma_1, \dots, \gamma_q)^\top$ is proportional 
%where $|J|$ is the absolute value of the Jacobian and $J = \prod_{i=1}^q\rho_i(1-\rho_i)$. 
In our analysis, for each model we ran two MCMC chains for 30,000 iterations each. Posterior inference was based upon 15,000 samples retained after adequate convergence was diagnosed. The MDAGAR model in the simulation examples was programmed in the S language as implemented in the \texttt{R} statistical computing environment. All other models were implemented using the \texttt{rjags} package available from CRAN \url{https://cran.r-project.org/web/packages/rjags/}.

\section{Supplementary Details in Simulation}
\label{sup_sim}
\subsection{WAIC, AMSE and D score}
For the simulation studies in Section~\ref{modelcom},  let $\bm{\theta} = \{\bm{\beta},  \bm{\sigma}, \bm{w}\}$. The likelihood of each data point $p(y_{ij} \mid \bm{\theta})= p\left(y_{ij} \mid \bm{x}_{ij}^{\top}\bm{\beta}_i + w_{ij}, 1/\sigma_{i}^{2}\right)$ is needed for calculating WAIC which is defined as 
\begin{align*}
WAIC = -2\left(\widehat{lpd} - \hat{p}_{WAIC}\right)\;,
\end{align*}
where $\widehat{lpd}$ is computed using posterior samples as the sum of log average predictive density i.e. $\sum_{i=1}^q\sum_{j=1}^{k}\log \left(\frac{1}{L}\sum_{\ell = 1}^Lp\left(y_{ij} \mid \bm{\theta}^{(\ell)}\right)\right)$, $\bm{\theta}^{(\ell)}$ for $\ell = 1, \dots, L$ being $L$ posterior samples of $\bm{\theta}$, and $\hat{p}_{WAIC}$ is the estimated effective number of parameters computed as 
\begin{align*}
\sum_{i=1}^q\sum_{j=1}^{k}V_{\ell = 1}^L\left(\log p\left(y_{ij} \mid \bm{\theta}^{(\ell)}\right)\right)
\end{align*}
with $V_{\ell = 1}^L\left(\log p\left(y_{ij} \mid \bm{\theta}^{(\ell)}\right)\right) =  \frac{1}{L-1}\sum_{\ell=1}^L\left[\log p\left(y_{ij} \mid \bm{\theta}^{(\ell)}\right) -\frac{1}{L}\sum_{\ell=1}^L\log p\left(y_{ij} \mid \bm{\theta}^{(\ell)}\right)\right]^2$.

Turning to the $D$ score, we draw replicates $y_{ij}$, $y_{\text{rep},ij}^{(\ell)} \sim N(\bm{x}_{ij}^{\top}\bm{\beta}_i^{(\ell)} + w_{ij}^{(\ell)}, 1/\sigma_{i}^{2(\ell)})$ and compute $D = G + P$. Here $G = \sum_{i=1}^q || \bm{y}_i - \bar{\bm{y}}_{\text{rep},i}||^2$ is a goodness-of-fit measure, where $\bar{\bm{y}}_{\text{rep},i}$ is the mean veactor with elements $\bar{y}_{\text{rep},ij} = \frac{1}{L}\sum_{\ell=1}^Ly_{\text{rep},ij}^{(\ell)}$ and $P = \sum_{i=1}^q\sum_{j=1}^k\sigma_{\text{rep},ij}^2$ is a summary of variance, where $\sigma_{\text{rep},ij}^2$ is the variance of $y_{\text{rep},ij}^{(\ell)}$ for $\ell = 1, \ldots, L$. 

For AMSE, we use $w_{ij}$ as the true value of each random effect and $\hat{w}_{ij}^{(n)}$ is the posterior mean of $w_{ij}$ for the data set $n$. The estimated AMSE is calculated as $\widehat{AMSE} = \frac{1}{Nqk}\sum_{n = 1}^N\sum_{i=1}^q\sum_{j=1}^k \left(\hat{w}_{ij}^{(n)} - w_{ij}\right)^2$ with associated Monte Carlo standard error estimate 
\begin{align*}
\widehat{SE}(\widehat{AMSE}) = \sqrt{\frac{1}{(Nqk)(Nqk-1)}\sum_{n = 1}^N\sum_{i=1}^q\sum_{j=1}^k \left[\left(\hat{w}_{ij}^{(n)} - w_{ij}\right)^2-\widehat{AMSE}\right]^2}.
\end{align*}

\subsection{Coverage Probability} \label{sup_cp}
\begin{figure}[H]
	
	\begin{subfigure}[t]{0.3\textwidth}
		\centering
		\includegraphics[scale=0.3]{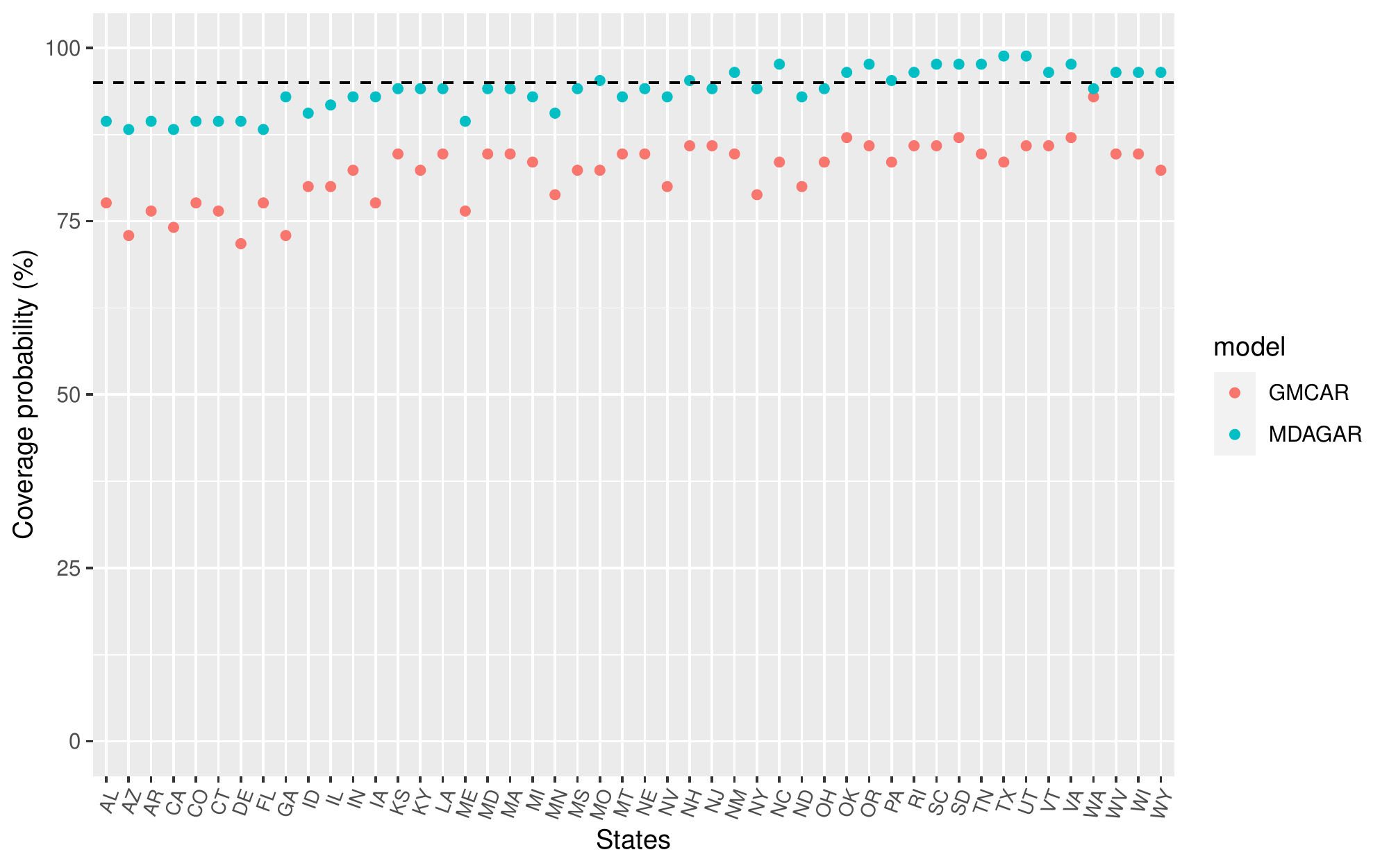}
		\caption{Low}
	\end{subfigure} 
	\begin{subfigure}[t]{0.3\textwidth}
		\centering
		\includegraphics[scale=0.3]{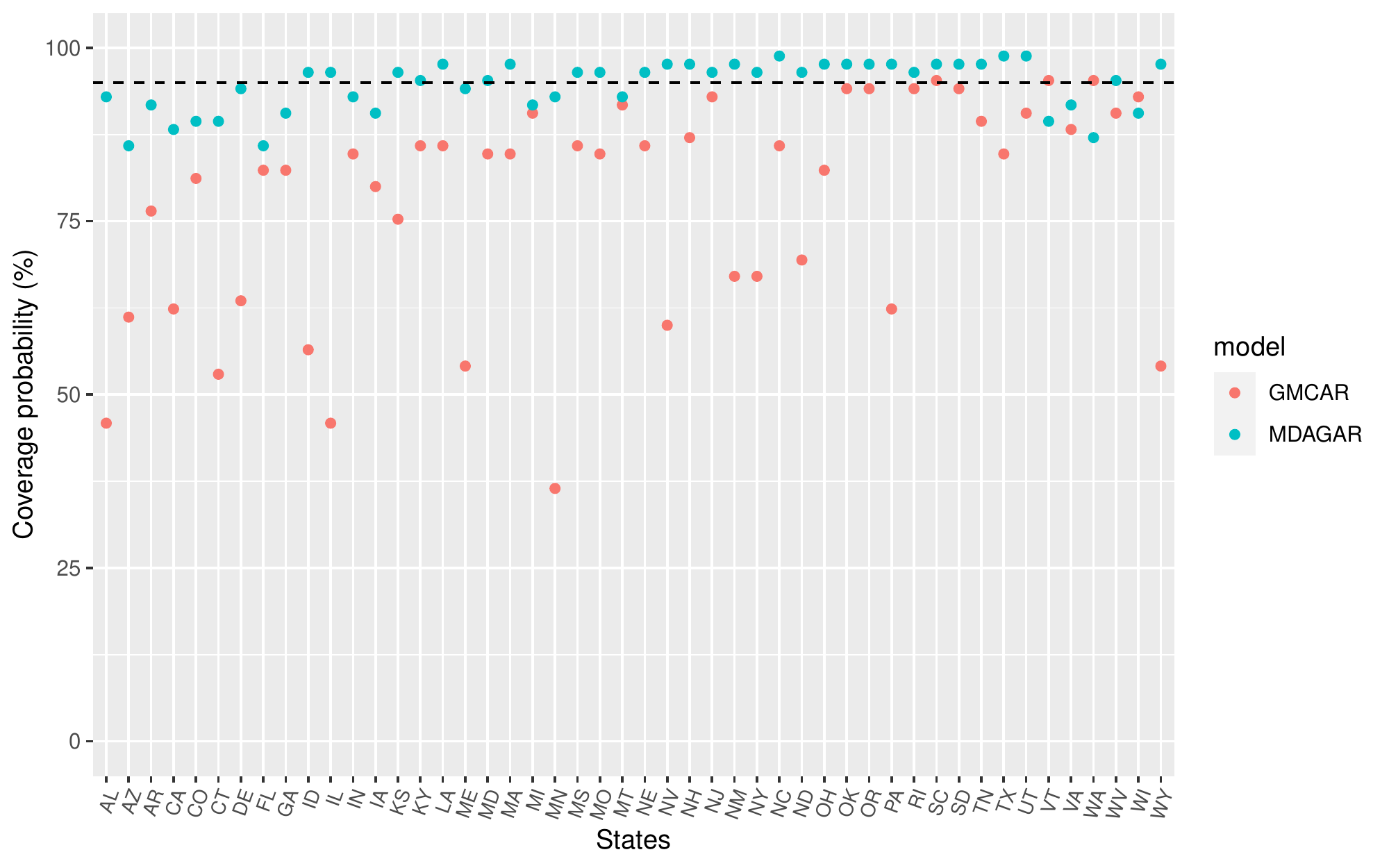}
		\caption{Medium}
	\end{subfigure} 
	\begin{subfigure}[t]{0.3\textwidth}
		\centering
		\includegraphics[scale=0.3]{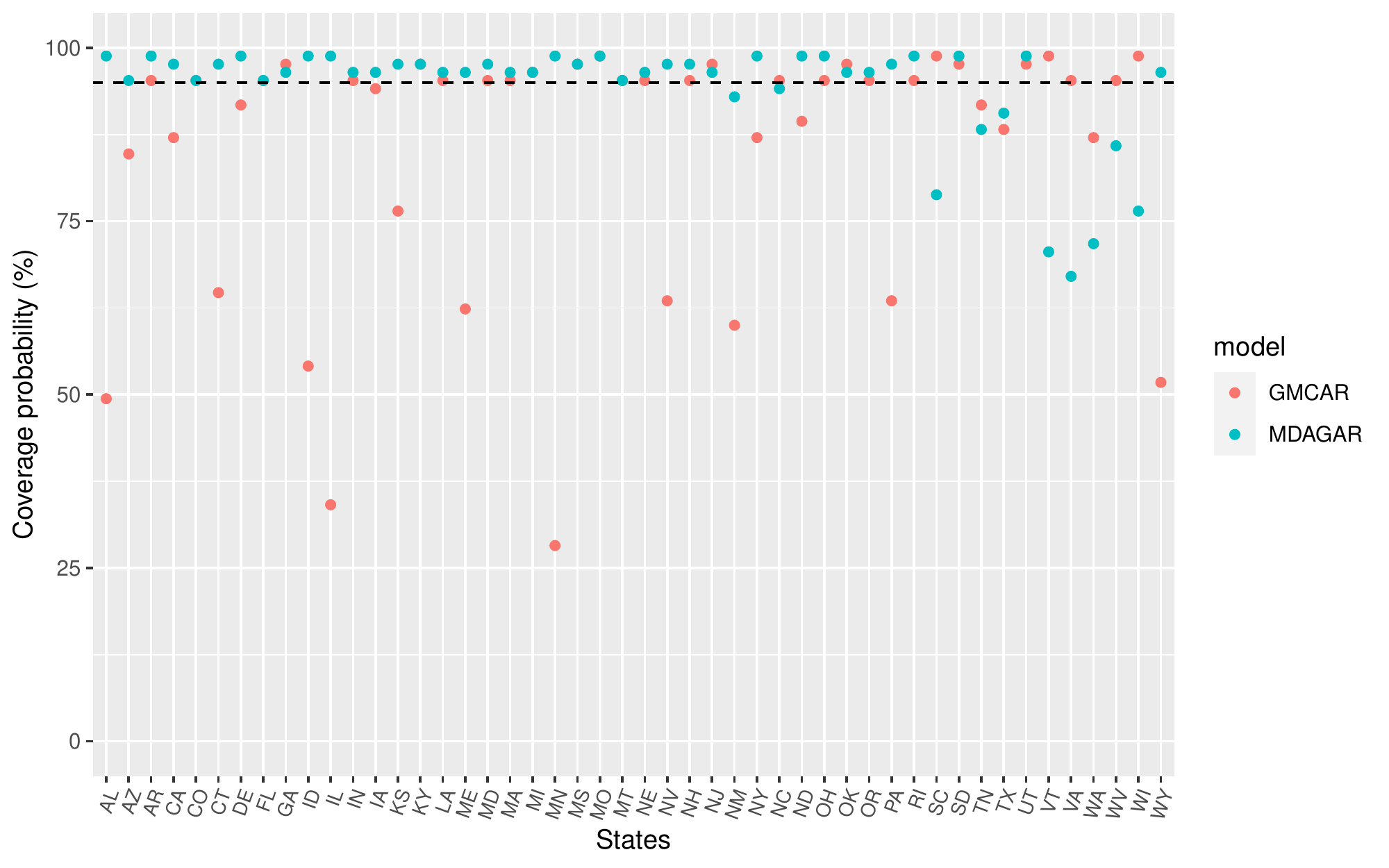}
		\caption{High}
	\end{subfigure}
	\caption{Coverage probability ($\%$) of $corr(w_{1j}, w_{2j})$, i.e. correlation between two diseases in each state, for MDAGAR (blue) and GMCAR (red).}\label{fig: cor}
	%		\end{adjustwidth}
\end{figure}

\begin{figure}[h]
	\centering
	\includegraphics[scale=0.4]{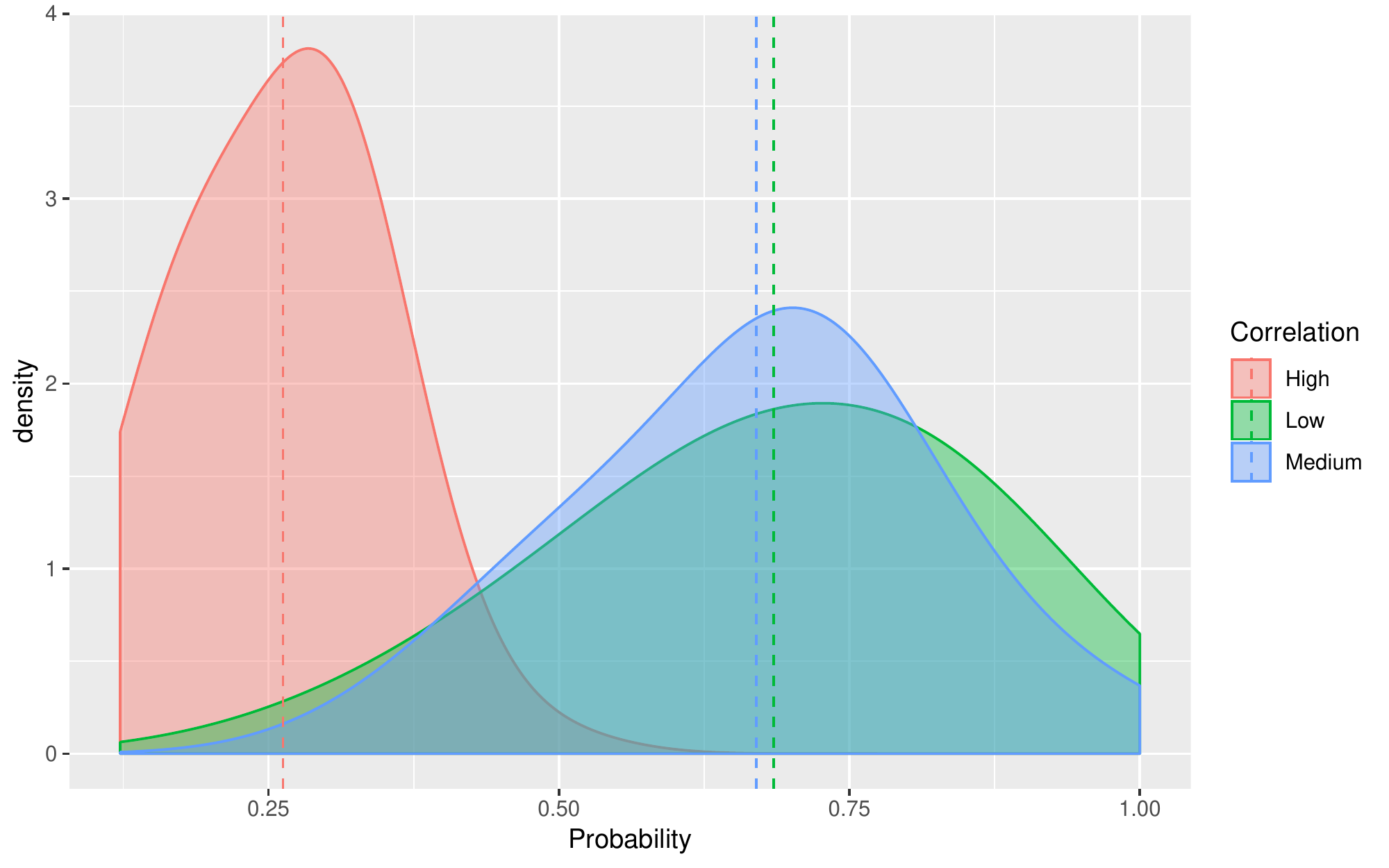}
	\caption{Density plots for probability that the KL-divergence between the MDAGAR and the true model is smaller than that between GMCAR and the true model with three levels of correlation for two diseases: low (purple), medium (green) and high (red)}\label{fig: KL}
\end{figure}

\begin{figure}
	\centering
	\begin{subfigure}{1\linewidth}
		\centering
		\includegraphics[width=80mm]{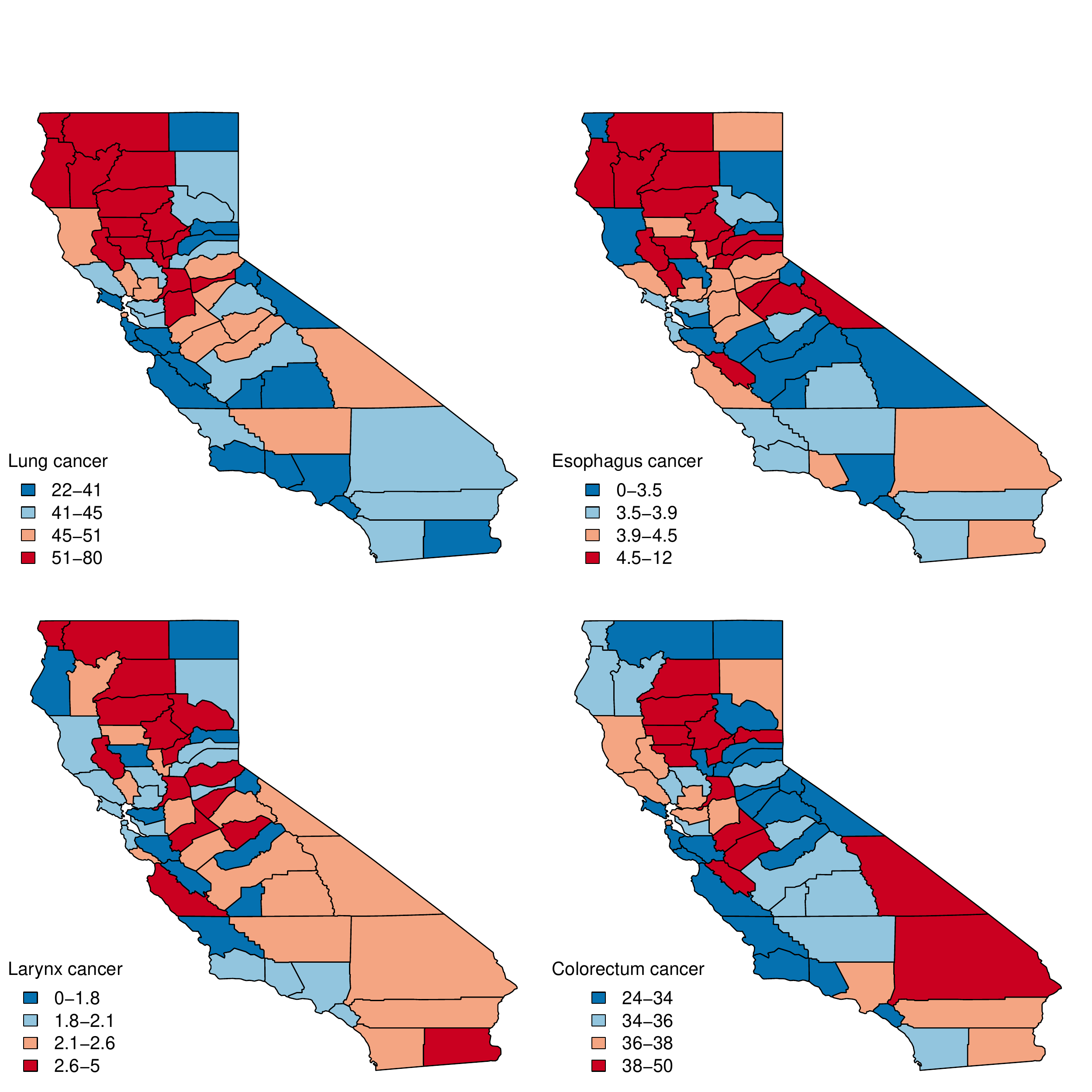}
		\caption{5-year average age-adjusted incidence rates per 100,000 population for lung, esophagus, larynx and colorectal cancer, 2012 - 2016}\label{fig: cancer_incidence_maps}
	\end{subfigure}
	\vfill
	\begin{subfigure}{1\linewidth}
		\centering
		\includegraphics[height= 70mm, width=120mm]{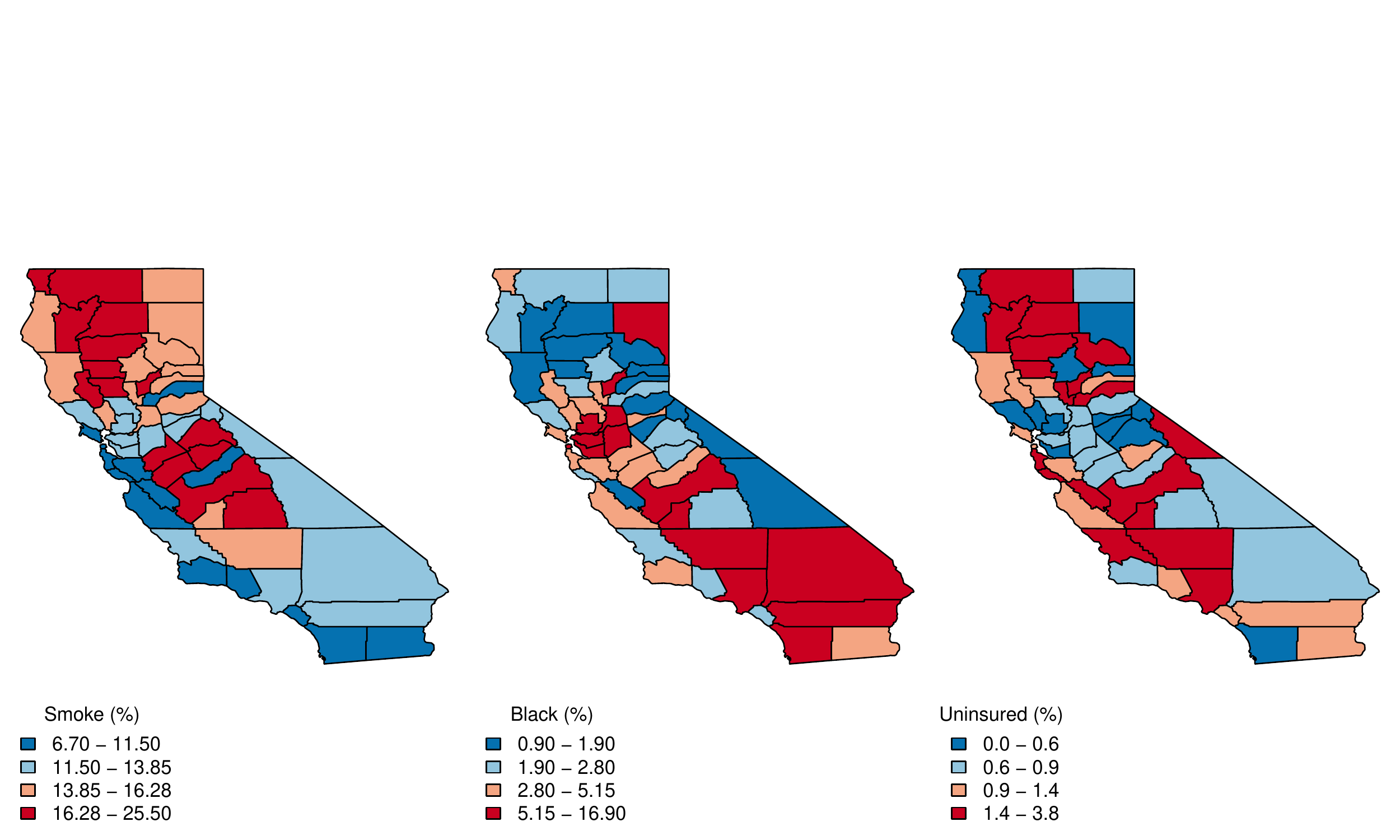}
		\caption{adult cigarette smoking rates (left), percentage of black residents (middle) and uninsured residents (right)}\label{fig: covariates}
	\end{subfigure}
	\caption{Maps of county-level raw data in California including (a) incidence rates for lung, esophagus, larynx and colorectal cancer and (b) important county-level covariates with significant effects: adult cigarette smoking rates, percentage of blacks and uninsured residents.}
\end{figure}

\begin{figure}[H]
	\centering
	\includegraphics[width=140mm]{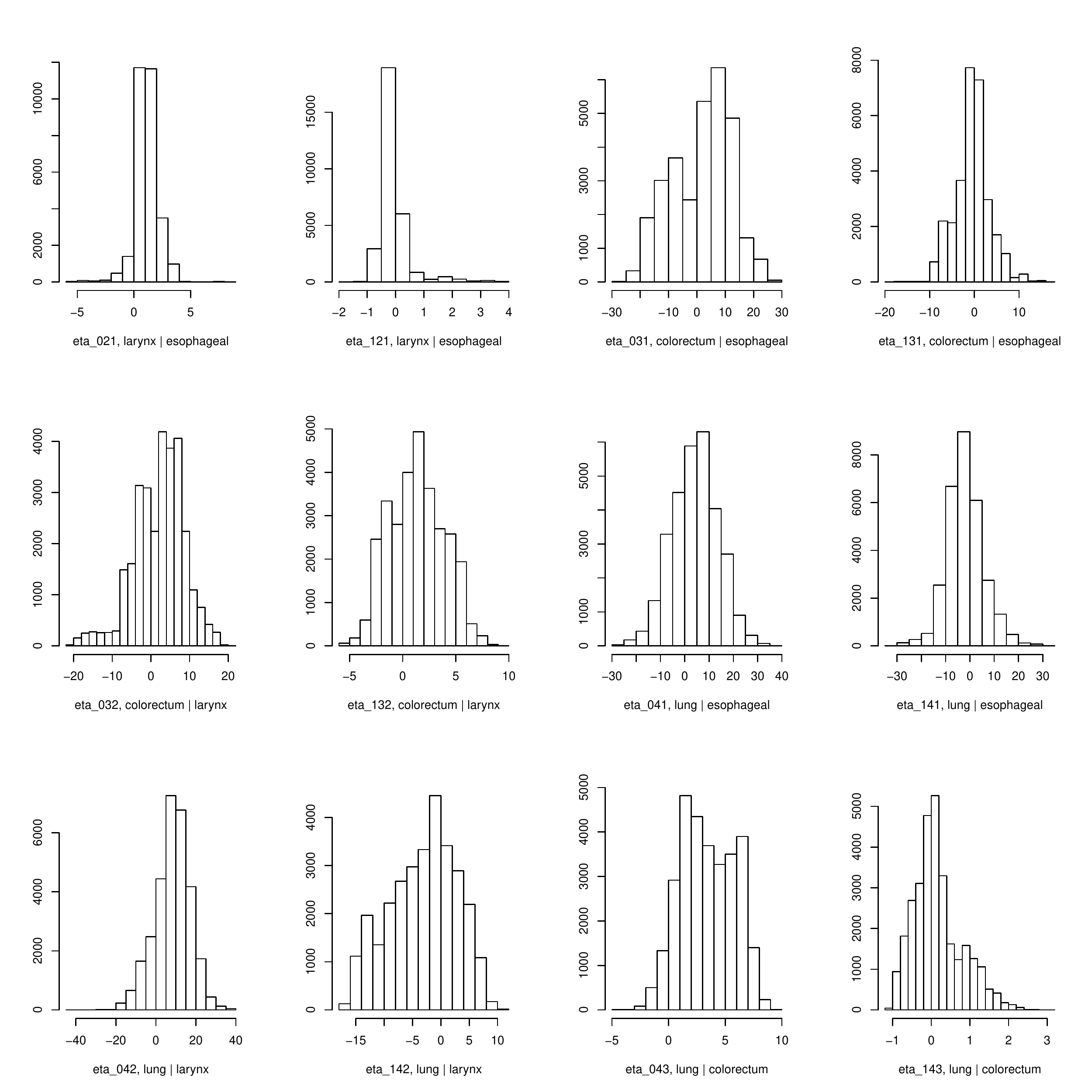}
	\caption{Posterior distributions of $\bm{\eta}$ for all pairs of cancers.}\label{fig: eta_final_new}
\end{figure}

\begin{figure}[H]
	\centering
	\includegraphics[width=\linewidth]{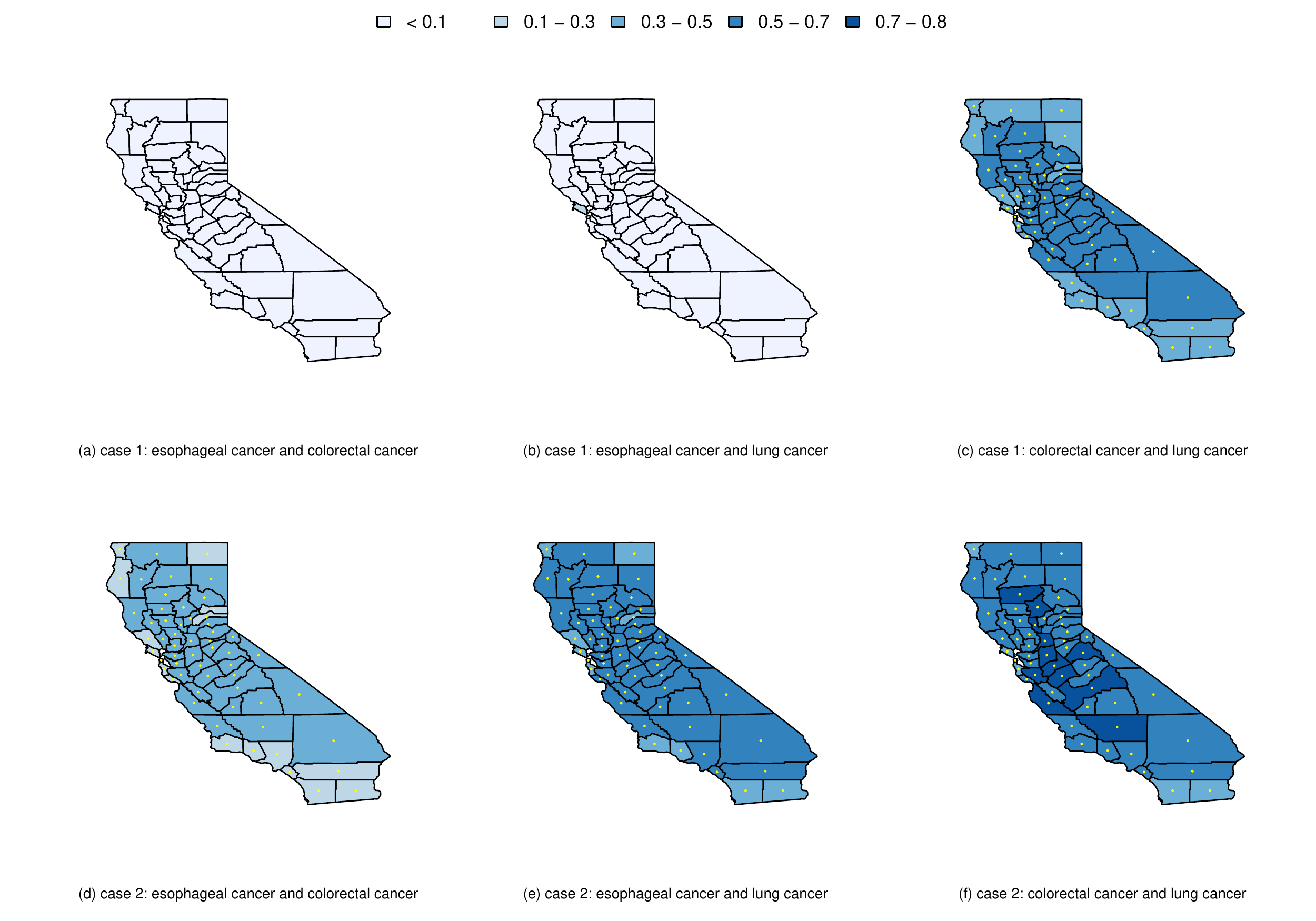}
	\caption{Estimated correlation between the incidence of pairwise cancers in each of 58 counties of California for Case 1 vs. Case 2: (a) case 1: esophageal and colorectal cancer, (b) case 1: esophageal and lung cancer, (c) case 1: colorectal and lung cancer, (d) case 2: esophageal and colorectal cancer, (e) case 2: esophageal and lung cancer, (f) case 2: colorectal and lung cancer. Maps (a)-(c) exihibit estimated correlations for Case 1, and (d) - (f) are for Case 2.  Yelllow points indicate significant correlations. Note: Maps for larynx cancer are not shown due to non-significant correlation with any of the other three cancers}\label{fig: correlation_cases}
\end{figure}

\begin{figure}[H]
	\centering
	\includegraphics[width=80mm]{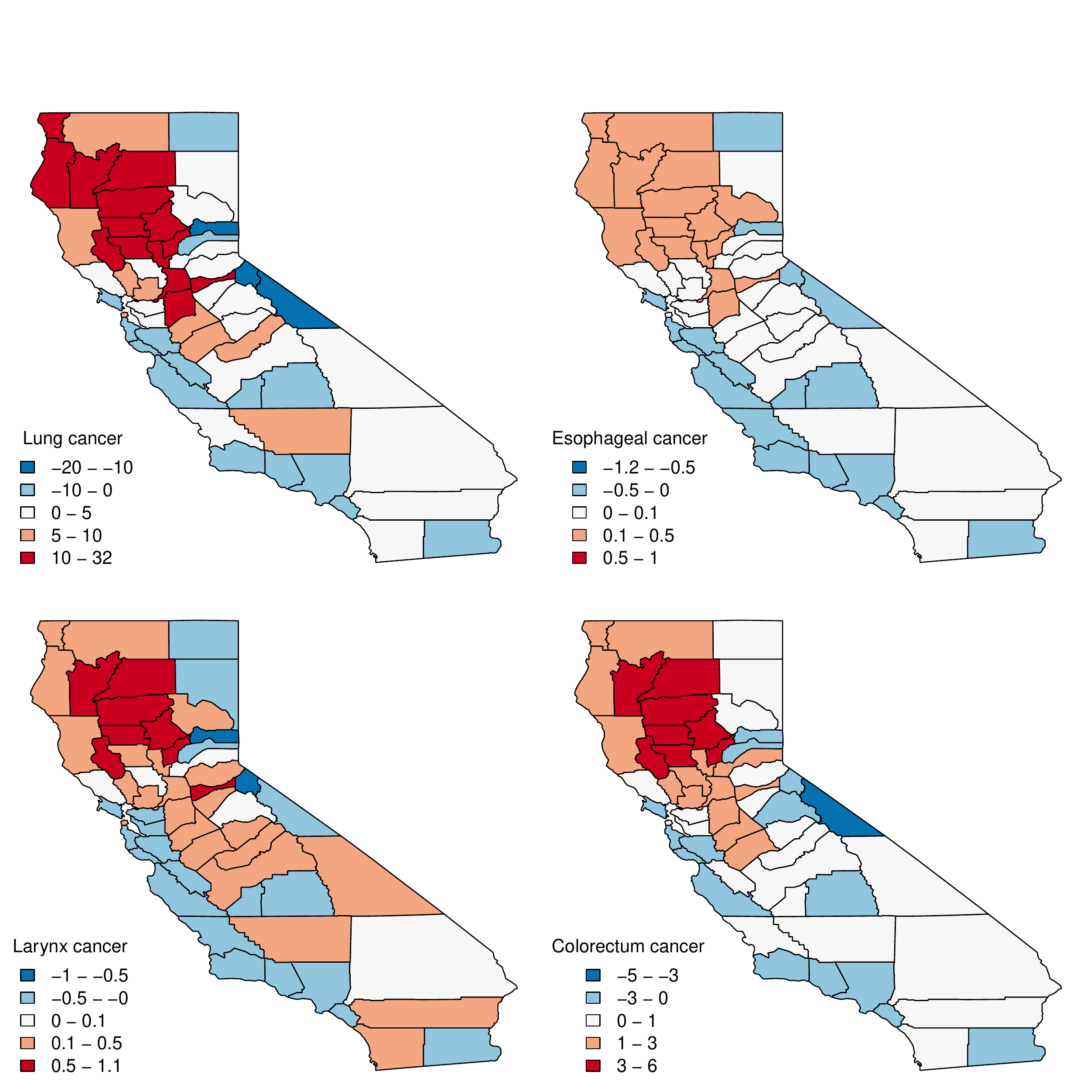}
	\caption{Maps of posterior mean spatial random effects (with no covariates) using the same order as $M_{10}$}\label{fig: random_nocovariates}
\end{figure}

For the simulation studies in Section~\ref{modelcom}, Table~\ref{tab:CP} shows the coverage probabilities (CP) defined as the mean coverage for a parameter by the $95\%$ credible intervals over 85 datasets. The MDAGAR offers satisfactory coverages for all parameters when correlations are low and medium, outperforming GMCAR, while GMCAR presents competitive performance with MDAGAR for high correlations. Figure~\ref{fig: cor} plots coverage probabilities of correlation between two diseases in the same region, given by $corr(w_{1j}, w_{2j}) = cov(w_{1j}, w_{2j})/(\surd var(w_{1j})\surd var(w_{2j}))$, for MDAGAR and GMCAR. Let $\bm{Q}(\rho_i)^{-1} = \{d_{ijj'}\}$, we obtain $cov(w_{1j}, w_{2j}) = \tau_1^{-1}(\eta_{021}d_{1jj} + \eta_{121}\sum_{j' \sim j}d_{1jj'})$, $var(w_{1j}) = \tau_1^{-1}d_{1jj}$ and
\begin{align*}
var(w_{2j}) = \tau_1^{-1}[\eta_{021}(\eta_{021}d_{1jj} + \eta_{121}\sum_{j'\sim j}d_{1jj'}) + \eta_{121}\sum_{j'\sim j}(\eta_{021}d_{1jj'}+\eta_{121}\sum_{j''\sim j}d_{1j''j'})] + \tau_2^{-1}d_{2jj}.
\end{align*}
The MDAGAR performs better in estimating disease correlations in the same region for all scenarios, especially for low and medium correlations with CPs at around $95\%$ in all states. 

%For high correlations, MDAGAR still offers satisfactory coverages in most states, except for states SC, TN, VT, VA, WA, WV and WI. The GMCAR model has better performance than the other two scenarios with most CPs at around $95\%$. However, it still shows small coverage for some states.

\begin{table}[H]
	\centering
	\caption{Coverage probability ($\%$) of parameters estimated from MDAGAR and GMCAR}\label{tab:CP}
	\vspace{1cm}
	\begin{tabular}{llcccccccc}
		\hline
		&&\multicolumn{8}{c}{Coverage probability ($\%$)}\\
		Correlation & Model& \multicolumn{1}{l}{$\eta_{021}$} & \multicolumn{1}{l}{$\eta_{121}$} & \multicolumn{1}{l}{$\rho_1$} & \multicolumn{1}{l}{$\rho_2$} & \multicolumn{1}{l}{$\tau_1$} & \multicolumn{1}{l}{$\tau_2$} & \multicolumn{1}{l}{$\sigma_{1}^2$} & \multicolumn{1}{l}{$\sigma_{2}^2$} \\
		\hline
		Low&MDAGAR&92.9  & 95.3  & 92.9  & 97.6  & 100   & 98.8  & 100   & 100 \\
		&GMCAR&92.9  & 80.0    & 84.7  & 95.3  & 95.3  & 100   & 83.5  & 100 \\
		
		Medium &MDAGAR&94.1  & 97.6  & 98.8  & 96.5  & 98.8  & 98.8  & 100   & 100 \\
		&GMCAR&85.9  & 77.6  & 69.4  & 98.8  & 61.2  & 98.8  & 84.7  & 98.8 \\
		High  & MDAGAR&92.9  & 94.1  & 95.3  & 54.1  & 96.5  & 98.8  & 78.8  & 100 \\
		&GMCAR&96.5  & 88.2  & 70.6  & 100   & 1.20   & 100   & 97.6  & 100 \\
		\hline
	\end{tabular}%
\end{table}

\section{Multiple Cancer Analysis from SEER for Case 2: Exclude smoking rates in covariates}
\label{sup_case2}
Excluding the risk factor adult cigarette smoking rates, we only include county attributes described in \ref{data} as covariates. Among $24$ models, model $M_{16}$ exhibits dominated best performance with a posteior probability of 0.999 and the corresponding conditional structure is $[\mbox{larynx}] \times [\mbox{esophagus}\given\mbox{larynx}] \times [\mbox{colorectal}\given \mbox{larynx}, \mbox{esophagus}] \times [\mbox{lung}\given \mbox{larynx}, \mbox{esophagus},\mbox{colorectal}]$. 

Table~\ref{tab: coeff} is a summary of the parameter estimates for each cancer. From $M_{16}$, we find that the regression slope for the percentage of blacks and unemployed residents are significantly positive for larynx and lung cancer respectively. The larynx cancer exhibits weaker spatial autocorrelation while the residual spatial autocorrelation for the other three cancers after accounting for preceding cancers are at moderate levels. For spatial precision $\tau_i$, larynx random effects still have the smallest variability while the conditional variability for colorectal and lung cancers are slightly larger.

\begin{table}[H]
	\centering
	\caption{Posterior means (95$\%$ credible intervals)for parameter estimated from $M_{16}$ for Case 2 (excluding somking rates in covariates).} \label{tab: coeff}
	\hspace*{-1.5cm}\begin{tabular}{ccccc}
		\hline
		Parameters& Larynx  & Esophageal & Colorectal & Lung \\
		\hline
		Intercept & 6.75 (-0.58, 14.00) & 11.14 (-1.70, 24.05) & 18.89 (-10.37, 48.12) & 24.18 (-22.71, 68.75) \\
		Young($\%$) & -0.09 (-0.20, 0.02) & -0.09 (-0.29, 0.11) & 0.27 (-0.19, 0.74) & 0.04 (-0.75, 0.86) \\
		Old($\%$) & -0.04 (-0.12, 0.04) & 0.00 (-0.15, 0.16) & 0.13 (-0.23, 0.49) & 0.15 (-0.49, 0.91) \\
		Edu($\%$)  & -0.02 (-0.08, 0.04) & -0.02 (-0.13, 0.09) & 0.12 (-0.13, 0.38) & -0.34 (-0.82, 0.15) \\
		Unemp($\%$)  & 0.04 (-0.03, 0.12) & 0.06 (-0.08, 0.20) & 0.10 (-0.26, 0.45) & \textbf{1.21 (0.55, 1.89)} \\
		Black($\%$)  & \textbf{0.15 (0.03, 0.27)} & 0.10 (-0.12, 0.32) & -0.20 (-0.75, 0.33) & 0.06 (-1.03, 1.13) \\
		Male($\%$) & -0.01 (-0.08, 0.07) & -0.07 (-0.21, 0.06) & 0.18 (-0.16, 0.52) & 0.01 (-0.59, 0.60) \\
		Uninsured($\%$)  & -0.07 (-0.19, 0.04) & -0.13 (-0.33, 0.07) & 0.10 (-0.37, 0.58) & 0.11 (-0.70, 0.95) \\
		Poverty($\%$) & 0.21 (-0.11, 0.53) & 0.40 (-0.20, 1.02) & 0.03 (-1.38, 1.45) & 0.84 (-2.14, 3.52) \\
		\hline
		$\rho_i$ & 0.25 (0.01, 0.91) & 0.49 (0.02, 0.97) & 0.43 (0.02, 0.94) & 0.50 (0.03, 0.98) \\
		$\tau_i$  & 44.04 (15.89, 90.23) & 24.55 (5.06, 61.33) & 18.25 (1.39, 51.15) & 19.68 (2.00, 55.07) \\
		$\sigma_{i}^2$ & 0.56 (0.37, 0.84) & 1.52 (0.88, 2.36) & 9.85 (6.48, 14.63) & 0.93 (0.18, 3.63) \\ 
		\hline
	\end{tabular}
\end{table}

%The posterior estimates for correlations between the incidence of pairwise cancers within each county in Figure~\ref{fig: correlation_cases} (d) - (f) show more significance and larger values compared with Case 1 in (a) - (c). The incidence of colorectum cancer and lung cancer are highly correlated in all counties with significant means at around $0.5 - 0.8$. The incidence of esophagus cancer also has significant correlation with colorectum and lung cancer separately in all counties, with means mostly at $0.3 - 0.5$ and $0.5-0.7$ respectively. Overall, several counties in the edge area exhibit much smaller correlation compared with other counties while counties in the middle with more neighbors tend to have higher correlation for the incidence of cancers. At last, similar to Case 1, the incidence of larynx cancer does not have significant correlation with any of the other three cancers and maps are not included.

\section{Supplementary Figures and Tables}
Supplementary figures and tables referenced in Section~\ref{modelcom}, \ref{data} and \ref{discussion} are shown below. 

%In Section~\ref{ca_result}, the histogram of posterior MCMC samples of $\bm{\eta}$'s for all pairs of cancers are shown in Figure~\ref{fig:eta_final}. Table~\ref{tab:coeff} shows 

\begin{table}[H]
	\centering
	\caption{The prosterior model probabilities for 24 models}\label{tab: post_prob}
	\begin{tabular}{cccccccc}
		\hline
		$p(M_1 \given \bm{y})$& $p(M_2 \given \bm{y})$ & $p(M_3 \given \bm{y})$& $p(M_4 \given \bm{y})$ & $p(M_5 \given \bm{y})$ & $p(M_6 \given \bm{y})$ & $p(M_7 \given \bm{y})$&$p(M_8 \given \bm{y})$\\
		\hline
		0.000 & 0.000 & 0.000 & 0.000 & 0.000 & 0.000 &0.000 &0.000 \\
		\hline\\
		\hline
		$p(M_9 \given \bm{y})$& $p(M_{10} \given \bm{y})$ & $p(M_{11} \given \bm{y})$& $p(M_{12} \given \bm{y})$ & $p(M_{13} \given \bm{y})$ & $p(M_{14} \given \bm{y})$ & $p(M_{15} \given \bm{y})$&$p(M_{16}\given \bm{y})$\\
		\hline
		0.000 & \textbf{0.577} & 0.000 & 0.000 & 0.000 & 0.000 &0.342 &0.079 \\
		\hline \\
		\hline
		$p(M_{17} \given \bm{y})$& $p(M_{18} \given \bm{y})$ & $p(M_{19} \given \bm{y})$& $p(M_{20} \given \bm{y})$ & $p(M_{21} \given \bm{y})$ & $p(M_{22} \given \bm{y})$ & $p(M_{23} \given \bm{y})$&$p(M_{24}\given \bm{y})$\\
		\hline
		0.000 & 0.000 & 0.000 & 0.000 & 0.000 & 0.000 &0.000 &0.002 \\
		\hline \\
	\end{tabular}
\end{table}
	
\end{document}